\documentclass[conference]{IEEEtran}
\IEEEoverridecommandlockouts

\usepackage{subcaption}

\usepackage{hyperref}
\usepackage{multirow}
\usepackage{xspace}
\usepackage{enumitem}
\usepackage{ulem}
\usepackage{booktabs} 
\usepackage{amssymb}  

\usepackage{graphicx}

\usepackage[table]{xcolor}

\usepackage{listings}
\usepackage{xcolor}
\definecolor{light-gray}{gray}{0.97}

\lstdefinestyle{promptstyle}{
    basicstyle=\rmfamily\footnotesize,
    backgroundcolor=\color{light-gray},
    keywordstyle=\color{blue},
    stringstyle=\color{red},
    commentstyle=\color{gray},
    showstringspaces=false,
    breaklines=true,
    tabsize=4
}

\definecolor{myblue}{RGB}{0, 114, 178}
\definecolor{myorange}{RGB}{230, 159, 0}
\definecolor{mygreen}{RGB}{3, 135, 9}
\definecolor{mypink}{RGB}{204, 121, 167}
\definecolor{mybrown}{RGB}{121, 79, 48}
\definecolor{myyellow}{RGB}{240, 228, 66}

\newcommand{\ignore}[1]{}

\renewcommand{\em}{\itshape}

\usepackage{xspace}
\usepackage{enumitem}
\usepackage{subcaption}

\newcommand{\mae}{MAE\xspace}
\newcommand{\mse}{MSE\xspace}

\newcommand{\cancer}{\texttt{cancer}\xspace}

\newcommand{\wine}{\texttt{wine}\xspace}
\newcommand{\housing}{\texttt{housing}\xspace}

\newcommand{\energy}{\texttt{energy}\xspace}

\newcommand{\qsar}{\texttt{biodegradation}\xspace}

\newcommand{\mcar}{MCAR\xspace}
\newcommand{\mar}{MAR\xspace}
\newcommand{\mnar}{MNAR\xspace}

\newcommand{\ece}{ECE\xspace}

\newcommand{\myparagraph}[1]{\vspace{1mm} \noindent {\em #1}}

\newcommand{\obs}{\text{obs}}
\newcommand{\mis}{\text{mis}}

\renewcommand{\emph}[1]{{\em #1}}

\newcommand{\nsamples}{\texttt{n-samples}\xspace}
\newcommand{\nruns}{\texttt{n-runs}\xspace}

\newcommand{\gain}{GAIN\xspace}
\newcommand{\gains}{GAIN-S\xspace}
\newcommand{\gainu}{GAIN-U\xspace}
\newcommand{\ot}{OT-Impute\xspace}
\newcommand{\miwae}{MIWAE\xspace}
\newcommand{\miwaeu}{MIWAE-U\xspace}
\newcommand{\miwaes}{MIWAE-S\xspace}
\newcommand{\mice}{MICE\xspace}
\newcommand{\tabcsdi}{TabCSDI\xspace}
\newcommand{\tabcsdis}{TabCSDI-S\xspace}
\newcommand{\softimpute}{SoftImpute\xspace}

\usepackage{amsmath,amssymb}

\newcommand{\Obs}{\text{obs}}

\newcommand{\bestcell}[1]{\cellcolor{black!20}{#1}}   
\newcommand{\secondcell}[1]{\cellcolor{black!10}{#1}}

\lstdefinelanguage{SQL}{
  morekeywords={SELECT,FROM,WHERE,AND,OR,NOT,JOIN,ON,AS,IN,EXISTS,GROUP, BY, HAVING},
  sensitive=false,
  morecomment=[l]--,
  morestring=[b]',
}
\lstdefinestyle{promptstyle}{
    basicstyle=\rmfamily\small,
    backgroundcolor=\color{light-gray},
    keywordstyle=\color{blue},
    stringstyle=\color{red},
    commentstyle=\color{gray},
    showstringspaces=false,
    breaklines=true,
    tabsize=4
}
\def\BibTeX{{\rm B\kern-.05em{\sc i\kern-.025em b}\kern-.08em
    T\kern-.1667em\lower.7ex\hbox{E}\kern-.125emX}}
\begin{document}


\title{Beyond Accuracy: An Empirical Study of Uncertainty Estimation in Imputation}

\author{
\IEEEauthorblockN{Zarin Tahia Hossain}
\IEEEauthorblockA{\textit{Department of Computer Science} \\
\textit{Western University}\\
London, Ontario, Canada \\
zarin.hossain@uwo.ca}
\and
\IEEEauthorblockN{Mostafa Milani}
\IEEEauthorblockA{\textit{Department of Computer Science} \\
\textit{Western University}\\
London, Ontario, Canada \\
mostafa.milani@uwo.ca}
}

\maketitle

\begin{abstract}
Handling missing data is a central challenge in data‐driven analysis. Modern imputation methods not only aim for accurate reconstruction but also differ in how they represent and quantify uncertainty. Yet, the reliability and calibration of these uncertainty estimates remain poorly understood. This paper presents a systematic empirical study of uncertainty in imputation, comparing representative methods from three major families: statistical (MICE, SoftImpute), distribution alignment (OT‐Impute), and deep generative (GAIN, MIWAE, TabCSDI). Experiments span multiple datasets, missingness mechanisms (MCAR, MAR, MNAR), and missingness rates. Uncertainty is estimated through three complementary routes: multi‐run variability, conditional sampling, and predictive‐distribution modeling and evaluated using calibration curves and the Expected Calibration Error (ECE). Results show that accuracy and calibration are often misaligned: models with high reconstruction accuracy do not necessarily yield reliable uncertainty. We analyze method specific trade‐offs among accuracy, calibration, and runtime, identify stable configurations, and offer guidelines for selecting uncertainty‐aware imputers in data cleaning and downstream machine learning pipelines. \end{abstract}


\section{Introduction}\label{sec:introduction}

Imputation is the process of estimating missing values in partially observed datasets, forming a foundational step in statistical analysis, machine learning pipelines, and data-centric scientific research. It underpins valid statistical inference, reliable model training, and robust decision systems in domains such as healthcare, finance, and the social sciences. A broad methodological landscape has emerged. Classical statistical approaches include deterministic rules and hot-deck procedures~\cite{andridge2010review}, likelihood-based formulations such as EM~\cite{dempster1977em} and Multiple Imputation (MI)~\cite{rubin2004multiple} with Bayesian MCMC variants~\cite{schafer1997analysis}. Machine learning methods frame imputation as prediction from observed features, including latent-variable models (probabilistic PCA)~\cite{tipping1999probabilistic,obermeyer2004bpca}, instance-based schemes (KNN)~\cite{troyanskaya2001missing,batista2002study}, chained equations (MICE)~\cite{buuren2011mice}, and ensemble models such as MissForest~\cite{stekhoven2012missforest}. Matrix and optimization-based approaches view it as low-rank recovery or regularized estimation, using nuclear-norm minimization~\cite{candes2009exact}, SoftImpute~\cite{mazumder2010spectral}, or matrix factorization~\cite{koren2009matrix}. Distributional alignment via optimal transport offers an alternative perspective, utilizing entropy-regularized Sinkhorn solvers to match observed and imputed distributions~\cite{cuturi2013sinkhorn}. Recent deep generative models leverage representation learning to capture complex dependencies, including denoising autoencoders (MIDA)~\cite{gondara2018mida}, variational methods (VAEAC, MIWAE)~\cite{ivanov2018vaeac,mattei2019miwae}, adversarial training (GAIN)~\cite{yoon2018gain}, and diffusion-based imputers for tabular data~\cite{tashiro2021csdi,kotelnikov2023tabddpm}.

While recovering plausible values is essential, an equally important question is how \emph{confident} we should be in those imputations. In modern ML, uncertainty is typically divided into aleatory and epistemic~\cite{huellermeier2021aleatoric,wimmer2023quantifying,smith2024rethinking,thawani2023chemistry}. Aleatory uncertainty reflects inherent variability in the data-generating process and remains even with unlimited data. Epistemic uncertainty arises from limited information or model misspecification and can, in principle, be reduced. Both appear in imputation: some missing values are genuinely unpredictable given the observed covariates (aleatory), while others are uncertain due to sparse features, strong modeling assumptions, or stochastic training (epistemic).

Reliable uncertainty is crucial in practice. It indicates how much trust to place in each imputed value and enables \emph{active imputation}, where high-uncertainty entries are prioritized for expert review or new data collection (e.g., remeasuring a patient’s blood pressure when the imputation seems unreliable). It also supports \emph{selective imputation} by flagging estimates that should not drive sensitive decisions, such as loan approvals. Under dataset shift, elevated uncertainty warns that the model is operating outside its training distribution, helping prevent overconfident, incorrect imputations. Uncertainty is not only a measure of reliability—it is a practical tool for guiding human oversight, managing risk, and improving data quality.

Several lines of work already produce uncertainty alongside imputations. Classical MI treats uncertainty as a first-class quantity by generating multiple randomized completions and combining within- and between-imputation variability~\cite{rubin2004multiple,schafer1997analysis,vanbuuren2018fid}. Likelihood-based and Bayesian implementations provide posterior draws for both parameters and missing data. Deep generative models yield samples from learned conditional distributions: VAE-based methods (e.g., VAEAC, MIWAE) produce stochastic imputations via latent-variable densities~\cite{ivanov2018vaeac,mattei2019miwae}; adversarial models such as GAIN support sampling through a generator–discriminator game~\cite{yoon2018gain}; and recent diffusion-based approaches instantiate flexible conditional samplers for complex missingness patterns~\cite{tashiro2021csdi}. In parallel, generic uncertainty tools—bootstrap resampling~\cite{efron1994bootstrap}, Bayesian deep learning approximations such as MC dropout~\cite{gal2016dropout}, and distribution-free conformal prediction~\cite{angelopoulos2023gentle} have been adapted to imputation pipelines without changing the core imputer.

However, despite this progress, we still lack a comprehensive understanding of \emph{how reliable} these uncertainties are across methods, datasets, and missingness regimes. Most empirical studies emphasize point accuracy (e.g., \mae or \mse) and report uncertainty only informally. As a result, it remains unclear whether the predictive distributions and intervals produced by these models are calibrated, whether nominal probabilities match empirical frequencies~\cite{gneiting2007scoring}. Consequently, practitioners have little guidance on the trade-offs between accuracy and uncertainty calibration. This gap is particularly evident across missingness mechanisms: Missing Completely at Random (\mcar), Missing at Random (\mar), and Missing Not at Random (\mnar), where the identifiability of conditional distributions and the influence of model assumptions differ substantially. It also persists across regimes of feature dimension, sample size, correlation structure, and missing rate, all of which can shift the calibration–accuracy balance.

This paper addresses these gaps with a systematic, uncertainty-centric evaluation of representative imputation approaches from three traditions: statistical and model-based methods (e.g., \mice), matrix/optimization-based methods (e.g., \softimpute, \ot) and deep generative models (e.g., \gain, \miwae, diffusion-based \tabcsdi). Our study is designed around three principles. First, we standardize \emph{how} uncertainty is extracted across methods, using either repeated runs with independent randomness, posterior or conditional sampling when available, or direct predictive distributions. Second, we evaluate uncertainty \emph{quality} using a calibration curve and the Expected Calibration Error (\ece), alongside accuracy metrics, so that calibration and point performance can be compared on equal footing~\cite{gneiting2007scoring}. Third, we evaluate across missingness mechanisms, missing rates, and diverse datasets to probe method behaviour under realistic variation in data structure and information loss.

Our findings reveal a consistent misalignment between accuracy and calibration. Methods that excel in point prediction are not necessarily those that provide well-calibrated uncertainty. For example, we observe that \mice, despite its simplicity, tends to maintain stable calibration across several datasets and missingness regimes, while \softimpute achieves competitive accuracy but exhibits poor calibration in many settings. Among deep generative approaches, \miwae often strikes a favourable balance between accuracy and calibration but at higher computational cost, reflecting the expense of latent-variable training and sampling. These results show that the choice of imputation strategy should depend on the requirements of the downstream task. When decisions rely on trustworthy uncertainty, such as in medical triage, selective automation, or human-in-the-loop review, it is preferable to use approaches that produce well-calibrated uncertainty, even if they sacrifice some point accuracy. In contrast, when both speed and high point-accuracy are the main priorities, approaches optimized for efficient and accurate imputation may be suitable, provided that their uncertainty estimates are calibrated afterward to ensure they remain reliable.

This work presents a unified framework and comprehensive empirical analysis of uncertainty in data imputation. By standardizing how uncertainty is extracted and evaluated under realistic missingness mechanisms, we clarify how different families of imputers trade off accuracy, calibration, and efficiency. The paper is structured as follows. Section~\ref{sec:related-work} reviews related work on imputation and uncertainty estimation. Section~\ref{sec:preliminaries} formalizes the imputation problem, missingness mechanisms, and representative methods. Section~\ref{sec:uncertanity} describes our experimental design, uncertainty extraction strategies, datasets, and evaluation metrics. Section~\ref{sec:experiments} presents empirical results on accuracy, calibration, and runtime, followed by key takeaways in Section~\ref{sec:analysis}. Finally, Section~\ref{sec:conclusion} concludes with a discussion of implications and future directions.

\section{Related Work}\label{sec:related-work}

Research on data imputation spans statistical heuristics, supervised learning, low-rank and factorization models, distribution-matching formulations, and deep generative methods~\cite{rubin1976inference,buuren2011mice}. Early baselines fill missing entries using variable-wise summaries or donor values (mean/median/mode and hot-deck). These are fast and easy to deploy but fail to preserve multivariate structure, often biasing correlations; they mainly serve as references for more principled approaches.

Supervised learning treats imputation as conditional prediction from observed features. Multiple Imputation (MI) adds stochasticity to generate multiple plausible completions and combine them for inference~\cite{rubin2004multiple}. A common variant, Multivariate Imputation by Chained Equations (\mice), models each variable given the others and injects randomness through coefficient or residual sampling~\cite{buuren2011mice}. Nonparametric and ensemble variants replace regressors with flexible learners: MissForest uses random forests~\cite{stekhoven2012missforest}, while KNN imputers average over local neighborhoods~\cite{troyanskaya2001missing}. These perform well for moderate dimensions and nonlinear relations and, when repeated, yield multiple imputations. 

A complementary line assumes a low-dimensional latent structure. Matrix completion methods recover missing entries from latent factors, with \softimpute performing iterative soft-thresholded SVD under a nuclear-norm penalty~\cite{mazumder2010spectral}. Probabilistic PCA and Bayesian variants sample posteriors for latent factors and missing values, enabling uncertainty-aware imputations~\cite{obermeyer2004bpca}. When correlations are captured by a few components, these methods achieve high accuracy with modest computation. Beyond low rank, distribution alignment casts imputation as matching empirical distributions of observed and completed data. Optimal Transport provides a geometric formulation solved via entropy-regularized Sinkhorn iterations~\cite{cuturi2013sinkhorn}; recent OT-based imputers align distributions while controlling reconstruction cost and preserving structure~\cite{muzellec2020missing}.

Deep generative models learn flexible conditional distributions for missing values. VAE-based methods reconstruct masked inputs through encoder–decoder architectures; \miwae improves training via importance weighting, and VAEAC supports arbitrary missingness patterns~\cite{mattei2019miwae,ivanov2018vaeac}. Adversarial models such as \gain rely on a discriminator to distinguish observed from imputed entries, capturing complex nonlinear dependencies~\cite{yoon2018gain}. Diffusion models offer another route by reversing a noise process, yielding high-quality, multimodal imputations for tabular data (e.g., TabDDPM)~\cite{kotelnikov2023tabddpm}. Although highly accurate, these neural approaches can be computationally demanding.

Uncertainty has deep roots in statistics and machine learning. Frequentist tools quantify sampling variability via confidence and prediction intervals~\cite{lehmann1998theory,casella2002statistical}, whereas Bayesian inference represents parameter and predictive uncertainty through posteriors~\cite{gelman2013bayesian}. In ML, concerns about overconfidence revived attention to calibration and to the distinction between aleatory and epistemic uncertainty~\cite{huellermeier2021aleatoric}. In imputation, MI explicitly targets uncertainty by combining within- and between-imputation variability~\cite{rubin2004multiple,vanbuuren2018fid}, and generic techniques like bootstrap, MC dropout, and conformal prediction provide uncertainty without modifying the imputer~\cite{efron1994bootstrap,gal2016dropout,angelopoulos2023gentle}. Recent work integrates uncertainty directly into modern imputers: conditional VAE frameworks estimate epistemic variance via dropout sampling, $\beta$-VAE variants trade sharpness for calibrated coverage, and retrieval-augmented Gaussian-process imputers yield posterior predictive variances that guide calibration and neighbour selection~\cite{hwang2022uncertainty,roskams2023leveraging,wang2024missing}. Empirical studies also show that calibrated imputation uncertainty enhances reliability in temporal clinical prediction and downstream ML pipelines~\cite{mulyadi2021uncertainty,cappiello2023effects,moslemi2024threshold}. 

Our work complements these efforts by providing a unified, uncertainty-centric evaluation of classical, optimization-based, and deep generative imputers. We standardize uncertainty extraction (via repeated runs, sampling, or predictive distributions) and assess reliability through calibration curves and Expected Calibration Error alongside point accuracy, clarifying trade-offs among accuracy, calibration, and computation across missingness mechanisms and data regimes.

Beyond imputation, probabilistic data-cleaning methods also model uncertainty explicitly. OTClean~\cite{OTClean2024} treats conditional-independence violations as distributional shifts and repairs them via optimal transport, while CurrentClean~\cite{CurrentClean2019,CurrentCleanICDE2019} addresses stale values by capturing temporal uncertainty. These methods use probabilistic structure to produce more reliable cleaned data, whereas we focus on calibrating uncertainty in missing-value imputation.

\section{Preliminaries}
\label{sec:preliminaries}

This section formalizes the imputation problem and reviews the details of the imputers evaluated in our study. 

\subsection{Imputation Problem and Missingness Mechanisms}

Let $X \in \mathbb{R}^{n \times d}$ be a data matrix with $n$ rows from an unknown distribution $\mathbb{P}_\theta$ and $d$ attributes. Missingness is represented by a binary mask $M \in \{0,1\}^{n \times d}$ with $M_{ij}=1$ if $X_{ij}$ is observed and $0$ otherwise. Using the Hadamard product $\,\odot\,$ and the all-ones matrix $\mathbf{1}_{n\times d}$,
\[
X^{\obs} \;=\; X \odot M, 
\qquad 
X^{\mis} \;=\; X \odot (\mathbf{1}_{n\times d} - M).
\]
We model the joint distribution of data and missingness as
\[
\mathbb{P}(X,M) \;=\; \mathbb{P}_\theta(X)\,\mathbb{P}_\phi(M\mid X),
\]
where $\mathbb{P}_\phi(M\mid X)$ is the missingness mechanism. The dataset-level imputation target is the conditional
\begin{align}
\mathbb{P}\!\left(X^{\mis}\mid X^{\obs},M\right) 
\;\propto\; 
\int \mathbb{P}_\theta(X)\,\mathbb{P}_\phi(M\mid X)\,\mathbb{P}(\theta)\, d\theta,
\label{eq:condi-impute}
\end{align}
from which samples yield plausible completions and conditional means provide point imputations. Focusing on a single record $x_i$ with mask $m_i$, while borrowing strength from the entire dataset through the posterior over $\theta$,
\begin{align}
&\mathbb{P}\left(x_i^{\mis} \mid x_i^{\obs}, m_i, X^{\obs}, M\right)=\nonumber\\
&\hspace{1cm}\int \mathbb{P}_\theta \left(x_i^{\mis} \mid x_i^{\obs} \right), \mathbb{P}(\theta \mid  X^{\obs}, M), d\theta.
    \label{eq:rec-cond}
\end{align}

\noindent Under row i.i.d.\ assumptions, conditioning on $\theta$ factorizes across records:
\[
\mathbb{P}_\theta\!\left(X^{\mis}\mid X^{\obs}\right)
\;=\;
\prod_{i=1}^n \mathbb{P}_\theta\!\left(x_i^{\mis}\mid x_i^{\obs}\right),
\]
so dependence across rows arises only from uncertainty in $\theta$ (or other shared latents) via $\mathbb{P}(\theta \mid X^{\obs},M)$.

We use Rubin’s taxonomy~\cite{rubin1976inference}. In \mcar, missingness is independent of the data, $\mathbb{P}_\phi(M\mid X)=\mathbb{P}_\phi(M)$; imputations can be accurate with well-calibrated uncertainty. In \mar, missingness depends only on observed values, $\mathbb{P}_\phi(M\mid X)=\mathbb{P}_\phi(M\mid X^{\obs})$; accurate imputation remains feasible if predictors of missingness are modeled, though uncertainty typically increases. In \mnar, $\mathbb{P}_\phi(M\mid X)\neq \mathbb{P}_\phi(M\mid X^{\obs})$; missingness depends on unobserved information (potentially the missing value itself), and both accuracy and calibration degrade unless the mechanism is explicitly modeled.

\subsection{Imputation Methods}  

We evaluate representative methods spanning classical regression-based multiple imputation, convex low-rank recovery, distribution matching via optimal transport, adversarial learning, variational inference, and diffusion-based conditional generation. Beyond point accuracy, we standardize how uncertainty is obtained (repeated runs, conditional/posterior sampling, or direct predictive distributions).

\myparagraph{1) \mice} is based on Multiple Imputation, MI,~\cite{rubin2004multiple} that treats missing values as random draws from their predictive distribution, producing several ($k$) completed datasets to reflect uncertainty due to missingness. Each dataset is analyzed separately, and results are pooled using Rubin’s rules, which combine within-imputation and between-imputation variances to yield valid statistical inference. While traditional MI relied on joint parametric models (e.g., multivariate normal), these approaches are infeasible for large, mixed-type datasets. Multiple Imputation by Chained Equations, \mice, addresses this by modeling each incomplete variable conditionally on others using regression models suitable for their data type (linear, logistic, Poisson, etc.). The algorithm iteratively imputes missing values until convergence, generating multiple completed datasets through stochastic draws.
\mice assumes data are \mar and performs best when strong auxiliary predictors are included. It allows users to enforce constraints and tailor models to individual variables. Although it does not always correspond to a coherent joint model and may be biased under \mnar conditions, \mice remains widely used because of its practicality, flexibility, and solid empirical performance in many real-world applications.

\myparagraph{2) \softimpute}~\cite{mazumder2010spectral} is a low-rank matrix completion method designed to efficiently recover missing entries in large data matrices. It assumes that the complete data can be well-approximated by a low-rank structure, where most variability is captured by a few latent factors (for example, user and item preferences in recommendation systems). 
The method formulates the imputation task as a convex optimization problem that balances reconstruction accuracy on observed entries with a nuclear-norm penalty on the estimated matrix. The nuclear norm serves as a convex surrogate for matrix rank, and the regularization parameter controls the effective dimensionality of the solution, similar to how the $\ell_1$ norm regularizes sparsity in Lasso regression.
Algorithmically, \softimpute iteratively fills in the missing values and applies soft-thresholded singular value decomposition (SVD) to update the estimate, shrinking singular values according to the regularization strength. Using warm starts and sparse matrix operations, the algorithm scales efficiently to large datasets. 
Although newer deep generative and diffusion-based methods often achieve higher accuracy and better uncertainty calibration, SoftImpute remains a foundational, interpretable, and computationally efficient baseline for large-scale imputation tasks based on low-rank modeling.

\myparagraph{3) \ot}~\cite{muzellec2020missing} performs imputation through distribution matching using optimal transport (OT). The method assumes that random subsets of the dataset should share the same distribution, encouraging imputations that preserve both local structure and global data geometry. 
It minimizes the Sinkhorn divergence between pairs of mini-batches sampled from the imputed matrix, which measures the transport cost required to align their empirical distributions. The Sinkhorn divergence is a differentiable, entropic-regularized approximation of the Wasserstein distance that can be efficiently optimized using matrix scaling algorithms.
In practice, missing entries are initialized with noisy column means and updated iteratively via stochastic gradient descent on the Sinkhorn loss. This nonparametric approach optimizes imputed values directly without assuming a specific generative model. A parametric extension, trained with the same loss in a round-robin fashion, enables out-of-sample imputation, though our experiments focus on the nonparametric variant. 
Empirically, \ot achieves strong performance under \mcar, \mar, and even \mnar mechanisms, offering a flexible alternative to low-rank and deep generative imputers by directly enforcing distributional alignment rather than relying on structural assumptions.

\myparagraph{4) \gain}~\cite{yoon2018gain} adapts the generative adversarial network (GAN) framework for missing data imputation. It consists of a generator that proposes imputations for missing values and a discriminator that attempts to distinguish between observed and imputed entries. 
The generator is trained through an adversarial loss to fool the discriminator while maintaining reconstruction accuracy on observed values through an additional loss term. By sampling different noise vectors, \gain naturally produces multiple imputations for the same partially observed instance. 
Empirically, \gain achieves strong accuracy on mixed-type datasets and moderate to high missingness rates. However, it inherits typical GAN challenges such as instability, sensitivity to hyperparameters, and limited theoretical guarantees beyond \mcar. Moreover, while it generates diverse imputations, it lacks calibrated uncertainty estimates. To address this, we extended \gain to a heteroscedastic version (denoted by GAIN-U in Section~\ref{sec:experiments}) where the generator outputs both means and variances, enabling uncertainty quantification while retaining the adversarial objective. It replaces the reconstruction MSE with a Gaussian negative log-likelihood on observed coordinates, while retaining the adversarial term.

\myparagraph{\miwae}~\cite{mattei2019miwae} extends the importance-weighted autoencoder (IWAE)~\cite{burda2016importance} to perform variational inference directly on incomplete data under the \mar assumption. It models each record using latent variables drawn from a prior and reconstructs the observed features through a neural decoder. An encoder network parameterized by $\gamma$ provides an amortized approximation to the posterior, producing distributional parameters (e.g., mean and variance) from partially observed inputs.
Training maximizes a missing-data importance-weighted bound, a tighter version of the evidence lower bound (ELBO) that uses multiple importance samples to approximate the observed-data log likelihood. Only observed entries contribute to the loss, enabling the model to train on incomplete records without discarding data. The bound approaches the true likelihood as the number of samples increases. \miwae combines probabilistic a approache with deep-learning scalability. It is theoretically sound under \mar, easy to train, and produces both point and multiple imputations that capture uncertainty.

\myparagraph{5) \tabcsdi}~\cite{zheng2022diffusion} extends conditional score-based diffusion models to tabular data with mixed numerical and categorical features. The method learns to generate plausible values for the missing part $x_i^\mis$ conditioned on the observed part $x_i^\Obs$ through a self-supervised denoising objective. During training, some features are randomly masked and reconstructed from noisy versions, while observed values remain fixed to provide context. The model performs a standard forward noising process and trains a neural network to predict the injected noise, effectively learning the reverse diffusion dynamics that recover clean samples during inference. Architecturally, \tabcsdi uses a transformer-based encoder adapted for non-temporal data. Multiple imputations are obtained by sampling different reverse diffusion trajectories, though the process is computationally expensive due to many reverse steps.  Its main limitations include high inference cost, difficulty with very high-cardinality categoricals, lack of explicit handling of \mnar mechanisms, and limited uncertainty calibration. Nonetheless, \tabcsdi offers a powerful likelihood-free framework for coherent, distribution-aware imputation.


\section{Experimental Methodology for Imputation Uncertainty}
\label{sec:uncertanity}

The purpose of this study is to evaluate imputation methods not only by their accuracy but also by the reliability of the uncertainty they provide. Most prior work has treated imputations as point estimates, focusing on minimizing reconstruction error. Yet missing values rarely have a single correct completion, and variability across plausible imputations can be as important as mean accuracy. Without uncertainty estimates, users risk overconfidence in the filled data, which can mislead downstream analyses and decision-making. This motivates a systematic comparison of state-of-the-art imputation methods under a controlled experimental framework. We aim to assess how uncertainty is captured, how well it is calibrated, and what trade-offs arise between accuracy, runtime, and uncertainty quality. The broader goal is to establish imputation methods as tools that provide not only completed datasets but also trustworthy information about the confidence of those completions.

\subsection{Approaches to Computing Uncertainty}
\label{sec:uncertainty_approaches}

There are many ways to quantify uncertainty in predictive modeling, but three approaches have emerged as the most natural and widely used in the context of data imputation. These approaches are consistent with common practices in statistics, ML, and generative modeling, and they align with how uncertainty is typically approximated when the true posterior distribution is intractable. Each reflects a different point of entry for stochasticity and provides a complementary view on uncertainty.

\begin{enumerate}[leftmargin=12pt]
    \item \emph{Repeated model runs.} 
    This approach quantifies uncertainty by running the imputer multiple times with different seeds, bootstrapped data, or initialization noise, and measuring the variability of the outputs. 
    We refer to this as the vanilla approach and use the plain method name to denote it (e.g., in the experiments, \gain refers to running the \gain imputer method multiple times).  
    The idea mirrors MI in classical missing-data analysis. The main challenge is computational cost: repeated training can be expensive for complex models such as GANs, VAEs, or diffusion models. 
    Another practical issue is deciding how many runs are sufficient to obtain stable variance estimates. Too few runs may underestimate epistemic variability, while too many runs can be prohibitively expensive.

    \item \emph{Sampling from the conditional distribution.} 
    Probabilistic generative models allow multiple imputations to be drawn from a fixed, trained model. 
    This probes aleatoric uncertainty, as the parameters are held fixed and randomness enters through the latent space or injected noise. 
    Compared to repeated runs, this approach is more efficient because training is done only once. 
    However, it requires models that explicitly support conditional sampling (e.g., \gain, \miwae, diffusion-based imputers). 
    The technical challenge is deciding how many samples to draw: too few gives noisy estimates of variability, while too many increases the runtime without much benefit. 
    For a model \texttt{X}, we use \texttt{X-S} to denote the sampling-based variant.

    \item \emph{Predictive distribution modeling.} 
    Some models are trained to output not only point estimates but also parameters of a predictive distribution (e.g., mean and variance for a Gaussian likelihood). 
    This approach provides per-cell parametric uncertainty directly as part of the model output. 
    Its main challenge lies in model design and training: the likelihood must be specified correctly, and the loss function must encourage meaningful variance estimation. 
    Poorly specified likelihoods can lead to overconfident or underconfident uncertainty estimates. 
    Moreover, these models can be harder to train and tune, since variance parameters are more sensitive to optimization instabilities. 
    For a model \texttt{X}, we use \texttt{X-U} to denote the uncertainty-output variant.
\end{enumerate}

Together, these strategies approximate the posterior predictive distribution in Eq~\ref{eq:condi-impute}
with different trade-offs between computational efficiency, modeling flexibility, and uncertainty calibration quality.

Each imputation algorithm in our study implements one or more of these strategies according to its modeling structure:
\begin{itemize}[leftmargin=*]
    \item \emph{\mice:} Supports only repeated runs (multiple chains); no sampling or uncertainty-output variant.
    \item \emph{\ot:} Provides uncertainty from repeated optimization with different initializations or minibatch orders.
    \item \emph{\softimpute:} Deterministic by design; any minor stochasticity from randomized SVD is treated as pseudo-uncertainty.
    \item \emph{\gain:} Supports all three approaches: multi-run (\gain), conditional sampling (\gains), and heteroscedastic predictive modeling (\gainu).
    \item \emph{\miwae:} Likewise supports multi-run (\miwae), conditional sampling (\miwaes), and decoder-based predictive variances (\miwaeu).
    \item \emph{\tabcsdi:} Supports multi-run (\tabcsdi) and sampling through multiple reverse-diffusion trajectories (\tabcsdis); no explicit uncertainty-output variant.
\end{itemize}

Overall, our evaluation includes six vanilla imputers
(\mice, \ot, \softimpute, \gain, \miwae, \tabcsdi),
three sampling-based variants
(\gains, \miwaes, \tabcsdis),
and two uncertainty-output models
(\gainu, \miwaeu).
Each yields an estimated predictive distribution
$\widehat{\mathbb{P}}(x_{ij}^{\mis}\mid X^{\obs},M)$
for every missing cell—either empirically from multiple imputations or parametrically from model outputs—
which we evaluate in terms of accuracy and calibration.

\subsection{Benchmarking Data}
\label{sec:data}

We use five \emph{numerical tabular} datasets spanning sizes and dimensionalities (Table~\ref{tab:datasets}). All features are z-scored to stabilize training and ensure comparable error scales.
\begin{table}[h]
\centering
\begin{tabular}{lrr}
\toprule
\textbf{Dataset} & \textbf{\#Rec.} & \textbf{\#Attr.} \\
\midrule
\housing & 20{,}640 & 8 \\
\qsar         &1055 &41\\
\cancer  & 569      & 30 \\
\energy  & 768      & 8 \\
\wine    & 178      & 13 \\
\bottomrule
\end{tabular}
\caption{Dataset statistics. All attributes are numerical.}
\label{tab:datasets}
\end{table}

All datasets used in this study are complete and therefore treated as ground truth. We adopt a semi-synthetic setup in which we start with fully observed data and inject missingness according to controlled \mcar, \mar, and \mnar mechanisms. Because the original datasets contain no missing values, we know the true values of every masked entry, enabling direct evaluation of both imputation accuracy and uncertainty calibration. Artificially injecting missingness also gives us full control over which features are affected, the conditions under which values become missing, and the overall missing rate, allowing systematic experimentation while preserving the real data distribution.

We injected missing values synthetically according to the three mechanisms (\mcar, \mar, \mnar). For each dataset, we fixed a target missingness rate (e.g., 10–15\%) and masked the corresponding number of entries. Under \mcar, cells were sampled uniformly at random across all rows and columns, with each position selected only once. Under \mar, we defined a dependency condition (such as another feature exceeding its mean), assigned higher masking probabilities to rows satisfying the condition, and sampled cells according to these probabilities until the desired rate was reached. Under \mnar, we specified a condition on the variable itself and assigned higher masking probabilities to cells meeting that condition, producing biased patterns in which certain values were more likely to be removed. Because sampling is probabilistic in \mar and \mnar, the final missing rate may differ slightly from the target. The normalized dataset prior to masking serves as the ground truth, and the corrupted dataset is used as input to the imputation algorithms. Full implementation details are available in our repository~\cite{zarin2025imputation_uncertainty}.

\subsection{Evaluation Measures}

We report runtime, {\em accuracy} via \mae, and {\em calibration} via calibration curves and \ece. Let $S\!=\!\{(i,j): M_{ij}=0\}$ denote masked cells. With ground truth $x_{ij}$ and imputed mean $\hat{x}_{ij}$,
\[
\mae=\frac{1}{|S|}\sum_{(i,j)\in S}\big|x_{ij}-\hat x_{ij}\big|.
\]
Lower \mae\ indicates better reconstruction on the same masked set. Calibration evaluates whether the model’s stated confidence matches empirical accuracy. For a grid of nominal coverage levels $q \in \{0,0.1,\ldots,1.0\}$, where each $q$ represents the confidence the model claims (e.g., $q{=}0.9$ corresponds to a 90\% prediction interval), we compute the model’s $q$-level intervals and measure the fraction of true values they contain. 
Plotting empirical coverage (y–axis) against $q$ (x–axis) yields the calibration curve: perfect calibration lies on $y=x$, while deviations indicate over- or under-confidence. We measure miscalibration with the expected calibration error
\[
\ece \;=\; \frac{1}{|Q|}\sum_{q\in Q} \big| \mathrm{Cov}(q)-q \big|,
\]
where $\mathrm{Cov}(q)$ is the observed coverage at level $q$. Lower values indicate better calibration. For continuous targets, we compute calibration using the CDF-based approach.

\section{Experimental Results}
\label{sec:experiments}

This chapter reports empirical results and analyzes uncertainty estimates across imputation methods. We first tune key parameters (Section~\ref{sec:tuning}), then summarize runtime (Section~\ref{sec:runtime}) and accuracy (Section~\ref{sec:acc}), and finally evaluate uncertainty and calibration (Section~\ref{sec:uncert-results}). All code, datasets, and additional figures are available in our repository~\cite{zarin2025imputation_uncertainty}.

\graphicspath{{figs/}}

\subsection{Tuning and Default Parameter Setting} \label{sec:tuning}
We tune these hyperparameters to balance accuracy and runtime:
\begin{itemize}[leftmargin=5mm]
    \item \mice: 20--80 iterations depending on dataset size.
    \item \ot: batch size 64--128; $\sim$300 iterations (small) to $\sim$500 (large).
    \item \softimpute: shrinkage $\lambda$ via CV on observed entries (log grid; \texttt{grid\_len}$=15$ small/medium, $=25$ large).
    \item \miwae: importance samples $K{=}10$; $1500$--$2500$ epochs.
    \item \gain: $1500$--$2500$ epochs; extra generator updates on high-$d$ data for stable training.
    \item \tabcsdi: epochs and reverse-diffusion steps capped for cost; e.g., \wine: \texttt{epochs}$=500$, \texttt{num\_steps}$=600$; \energy: \texttt{epochs}$=400$, \texttt{num\_steps}$=1000$.
\end{itemize} 

\begin{figure}
  \centering
  \begin{subfigure}{0.24\textwidth}
    \includegraphics[width=\linewidth]{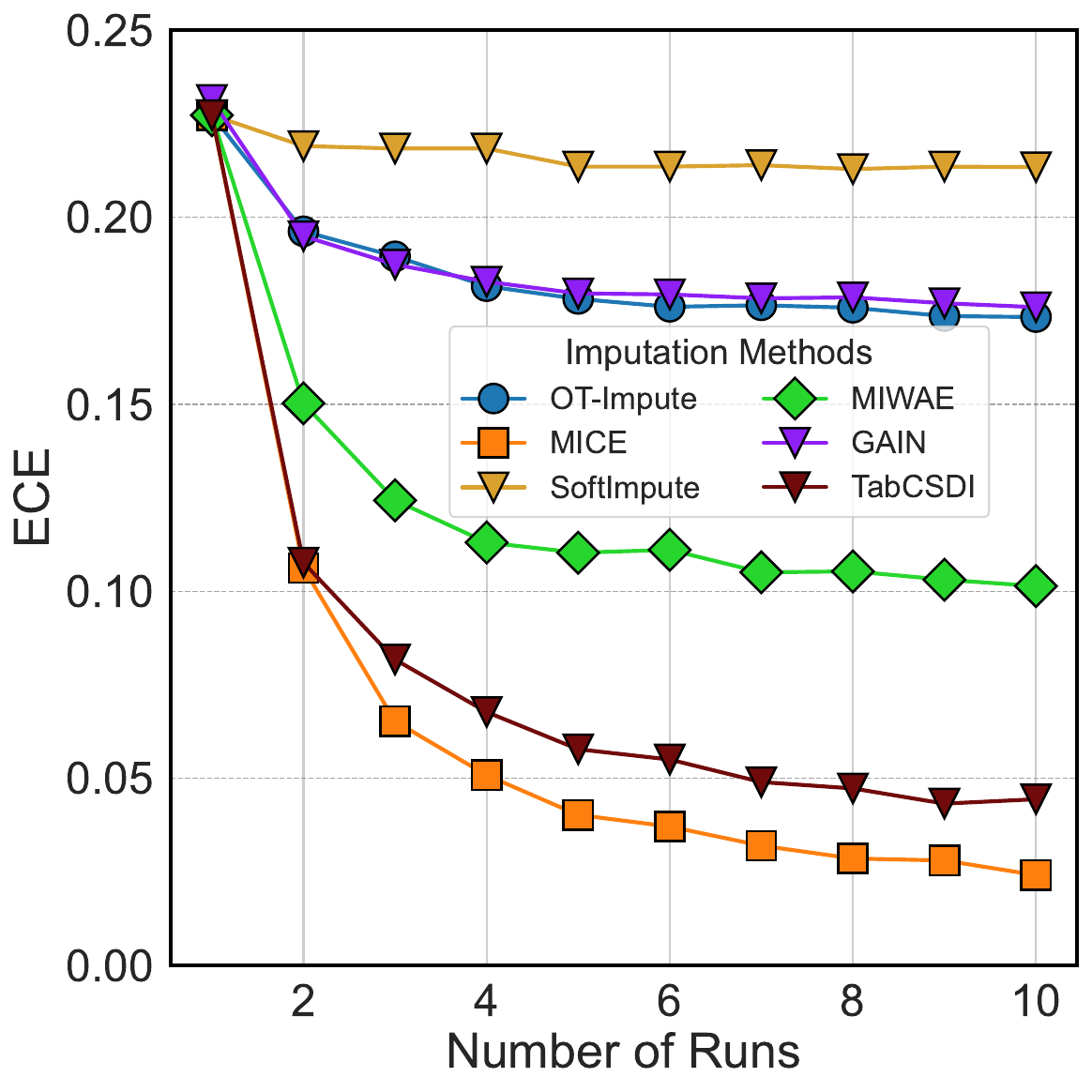}
    \caption{\wine}
    \label{fig:runs-mcar30}
  \end{subfigure}
  \begin{subfigure}{0.24\textwidth}
    \includegraphics[width=\linewidth]{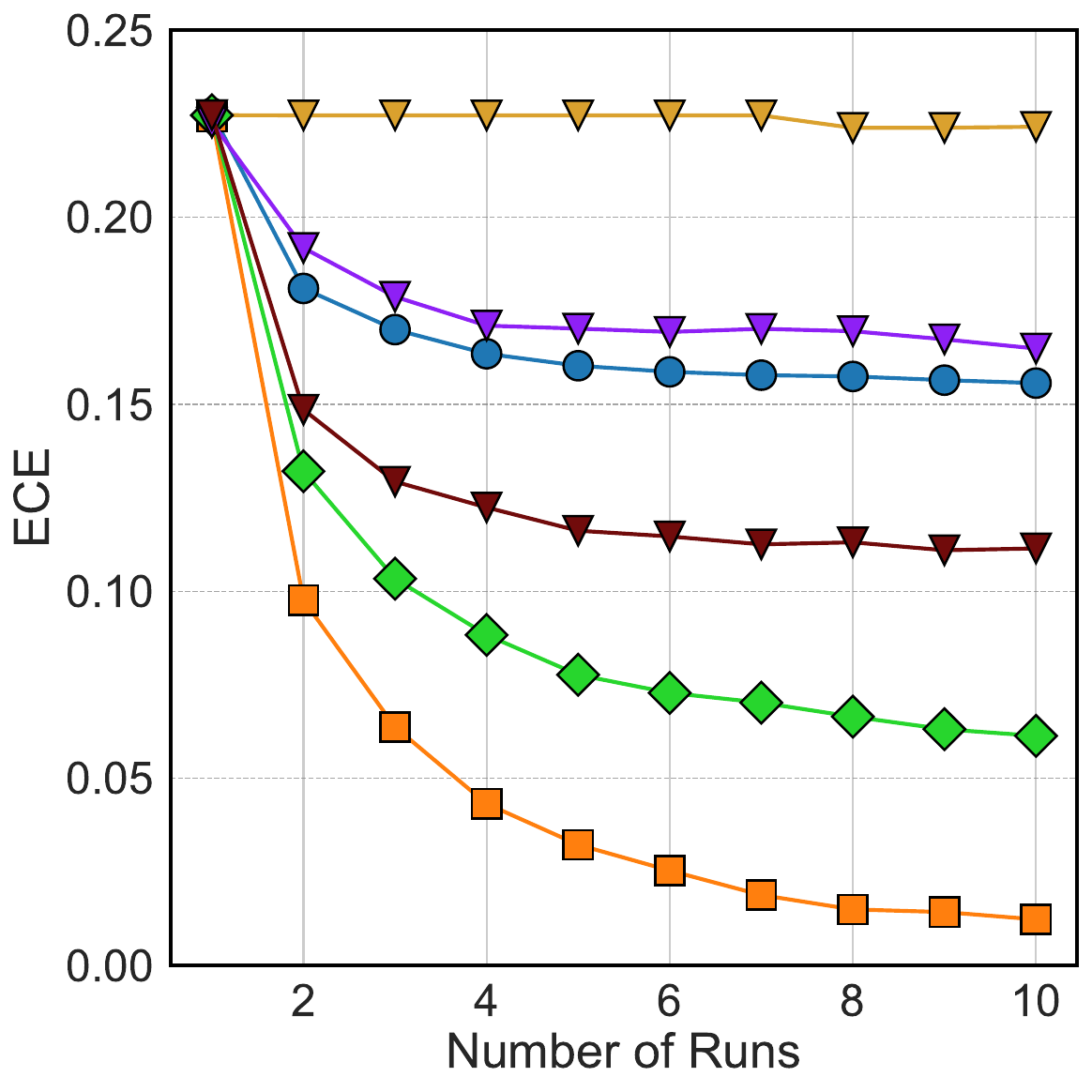}
    \caption{\energy}
    \label{fig:runs-mcar30-energy}
  \end{subfigure}
  \caption{\ece\ vs.\ \nruns\ at 30\% \mcar.}
  \label{fig:ece-runs}
\end{figure}

\begin{figure}
  \centering
  \begin{subfigure}{0.24\textwidth}
    \includegraphics[width=\linewidth]{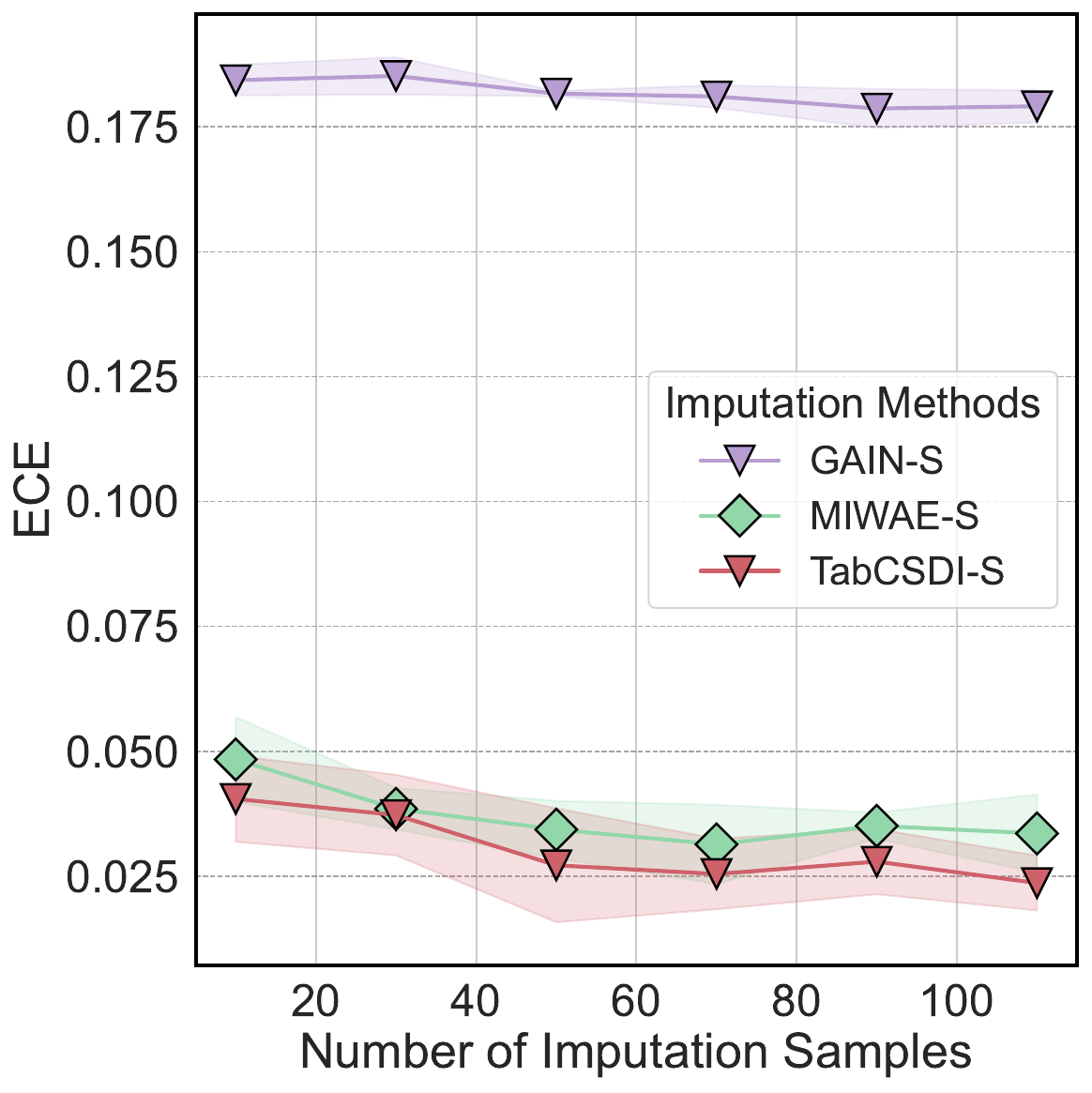}
    \caption{\wine}
    \label{fig:samp-mcar30-wine}
  \end{subfigure}
  \begin{subfigure}{0.24\textwidth}
    \includegraphics[width=\linewidth]{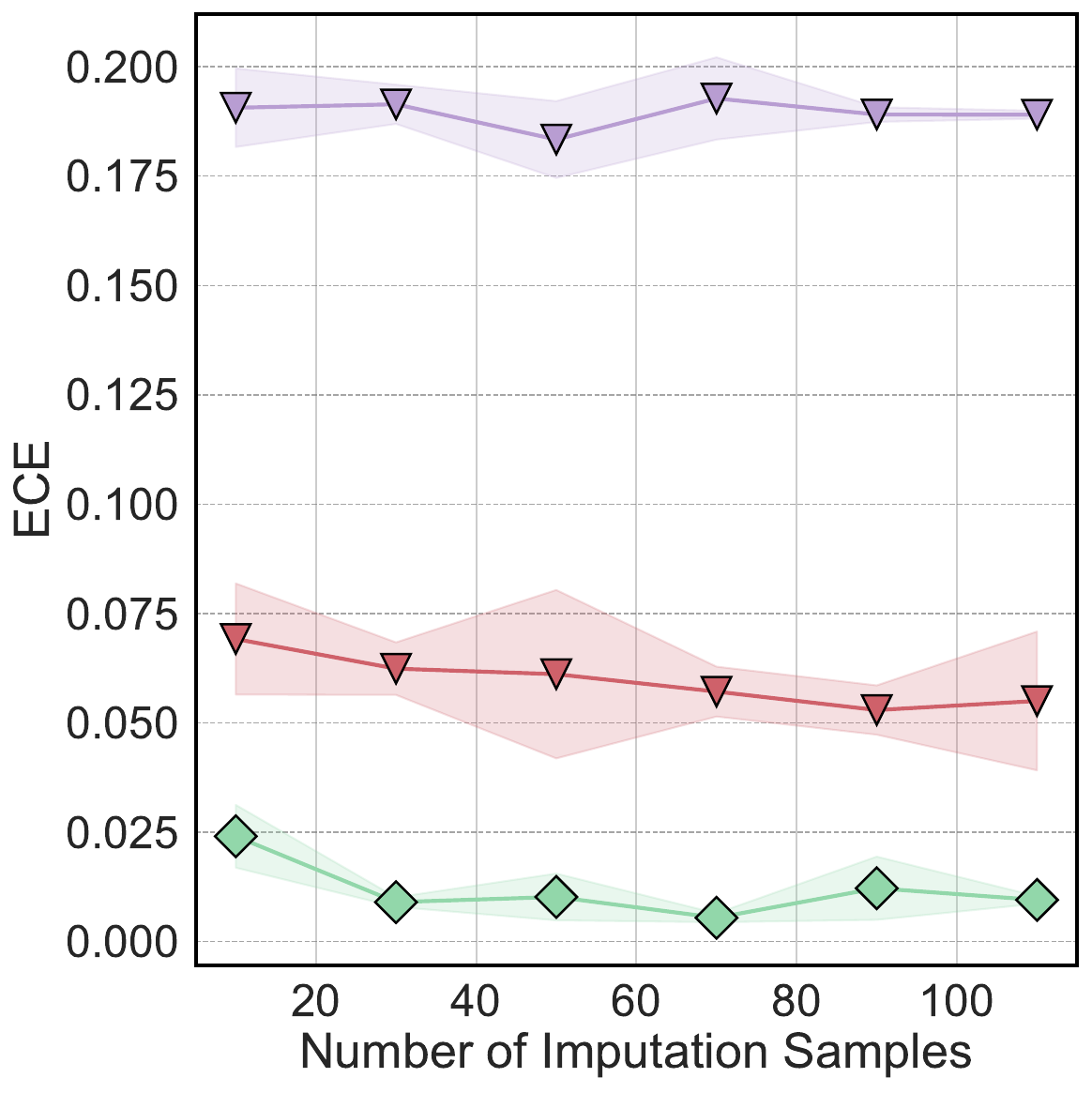}
    \caption{\energy}
    \label{fig:samp-mcar30-energy}
  \end{subfigure}
  \vspace{-5mm}
  \caption{\ece\ vs.\ number of samples at 30\% \mcar.}
  \label{fig:ece-sample}
\end{figure}

Figure~\ref{fig:ece-runs} shows that \ece\ steadily improves with increasing \nruns, but the gains plateau around five runs. 
Accordingly, we set \nruns$=5$ as the default for all multi-run experiments. 
For sampling-based variants (Figure~\ref{fig:ece-sample}), both \miwaes\ and \gains\ reach stable calibration by $\nsamples\!\approx\!20$, while \tabcsdis\ exhibits its best calibration performance between 50 to 70 samples.We therefore adopt $\nsamples{=}20$ for \miwaes\ and \gains, and $\nsamples{=}50$ for \tabcsdis.

\begin{figure}
  \centering
  \begin{subfigure}{0.24\textwidth}
    \includegraphics[width=\linewidth]{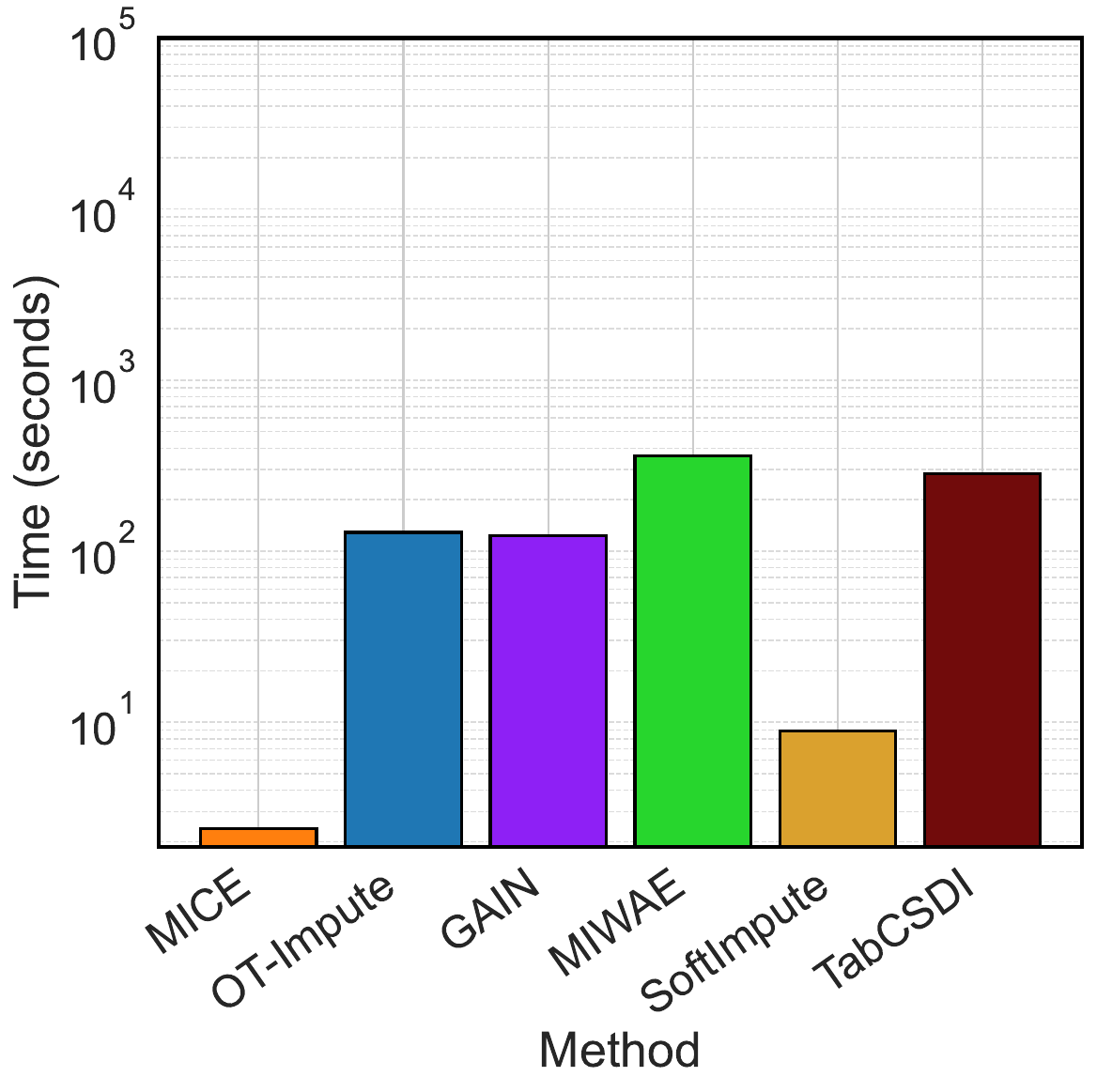}
    \caption{\wine}
    \label{fig:timebar-wine}
  \end{subfigure}\hfill
  \begin{subfigure}{0.24\textwidth}
    \includegraphics[width=\linewidth]{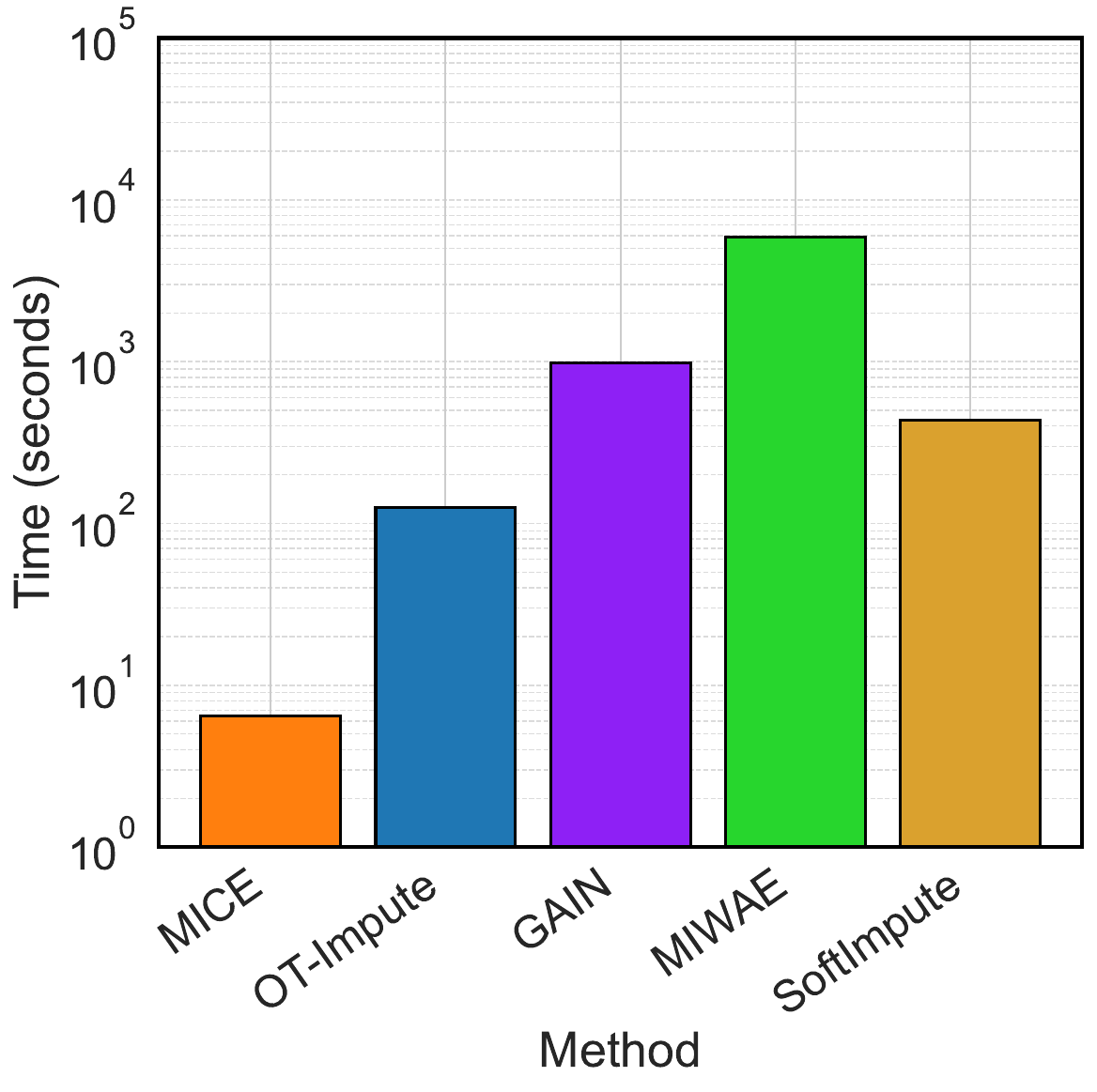}
    \caption{\housing}
    \label{fig:timebar-gas}
  \end{subfigure}
  \vspace{-2mm}
  \caption{Time per run at 30\% missingness vs Classical methods report total time; deep models report train$+$single imputation.}
  \label{fig:timebar-all}
\end{figure}

\begin{figure}[h]
  \centering
  \begin{subfigure}{0.24\textwidth}
    \includegraphics[width=\linewidth]{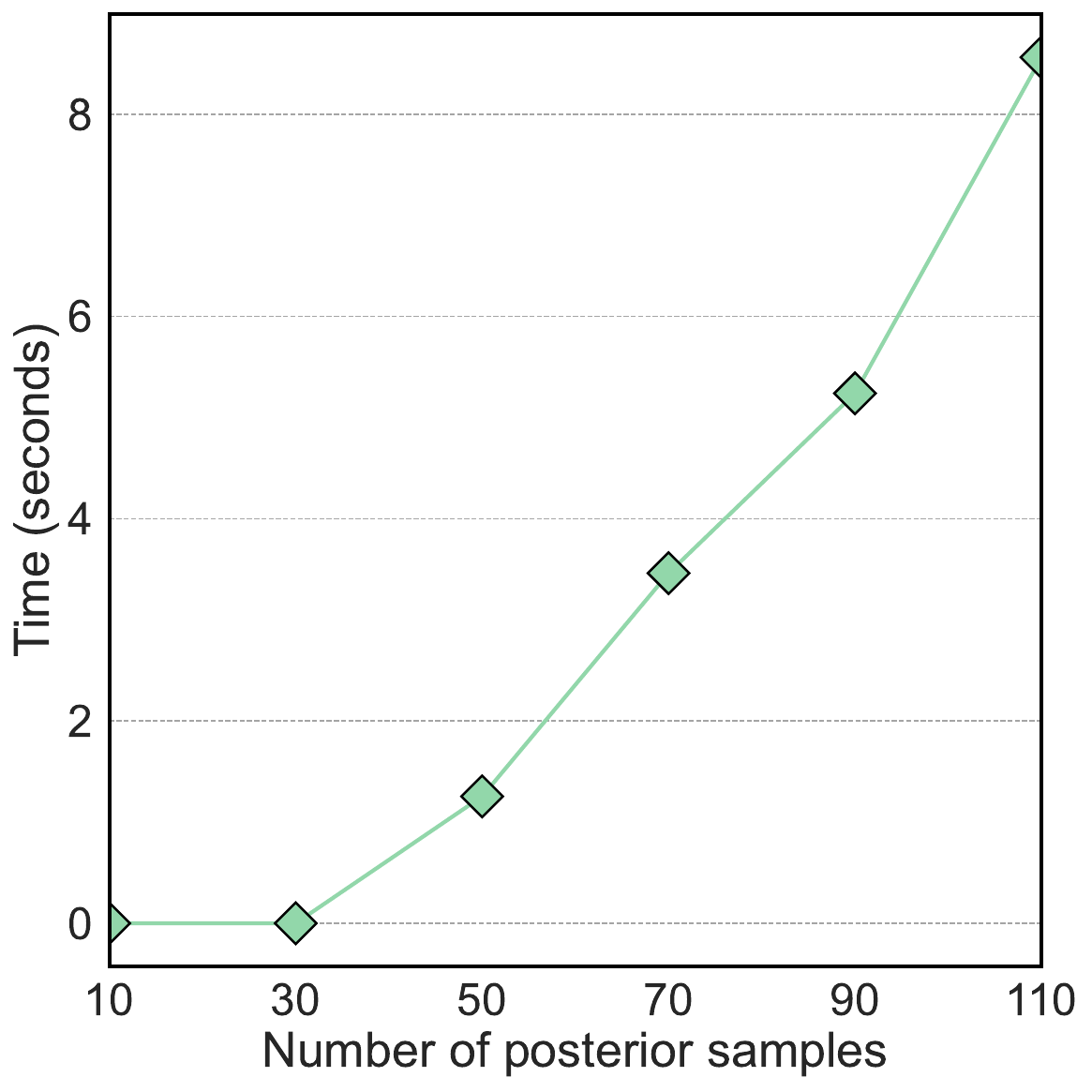}
    \caption{\miwaes}
  \end{subfigure}
  \begin{subfigure}{0.24\textwidth}
    \includegraphics[width=\linewidth]{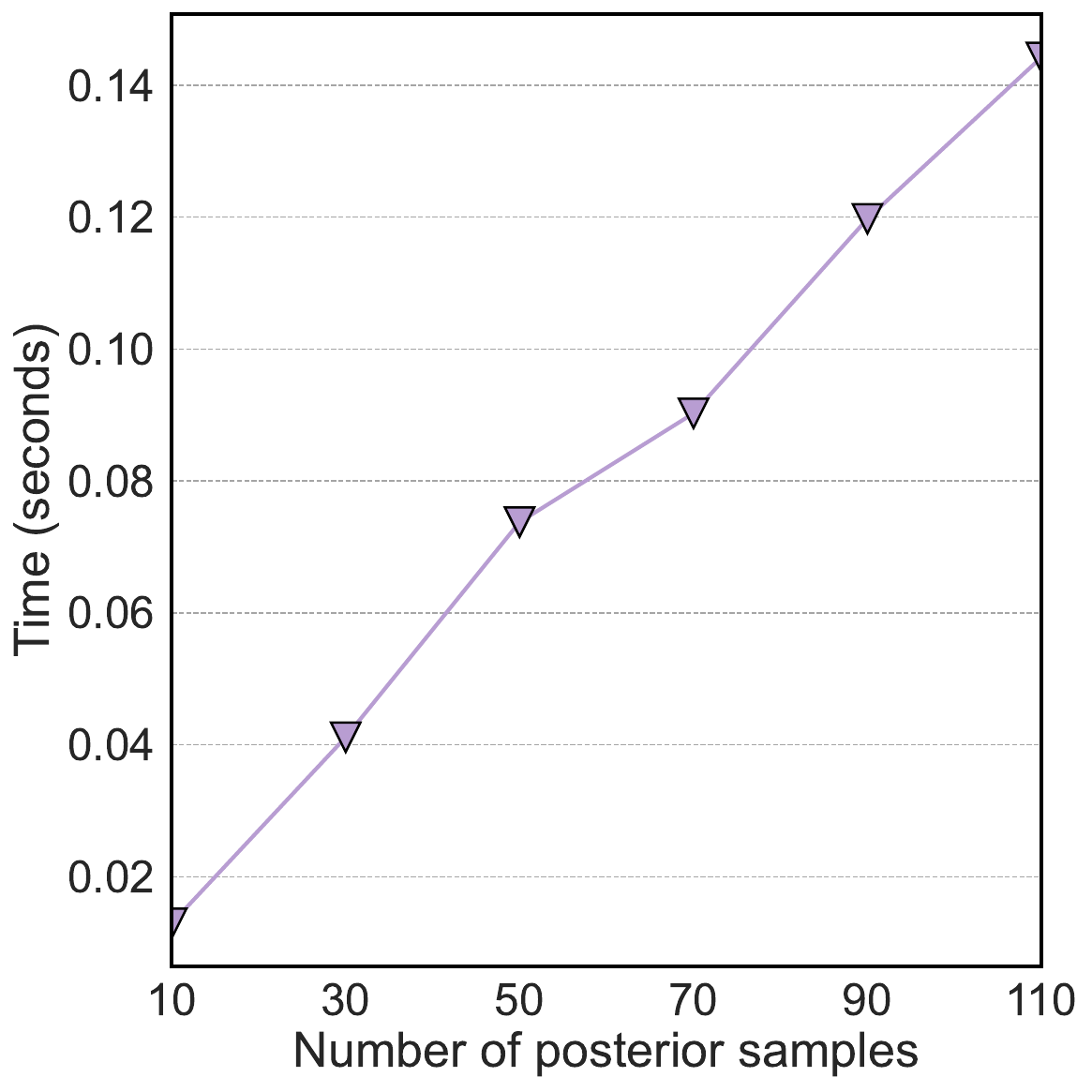}
    \caption{\gains}
  \end{subfigure}
  \vspace{-5mm}
  \caption{Runtime vs.\ \nsamples\ at 30\% \mcar\ in \wine.}
  \label{fig:sampling-time-S-variants}
\end{figure}


\begin{center}
\graphicspath{{figs/}}

\begin{minipage}[t]{0.48\columnwidth}
  \centering
\includegraphics[width=\linewidth]{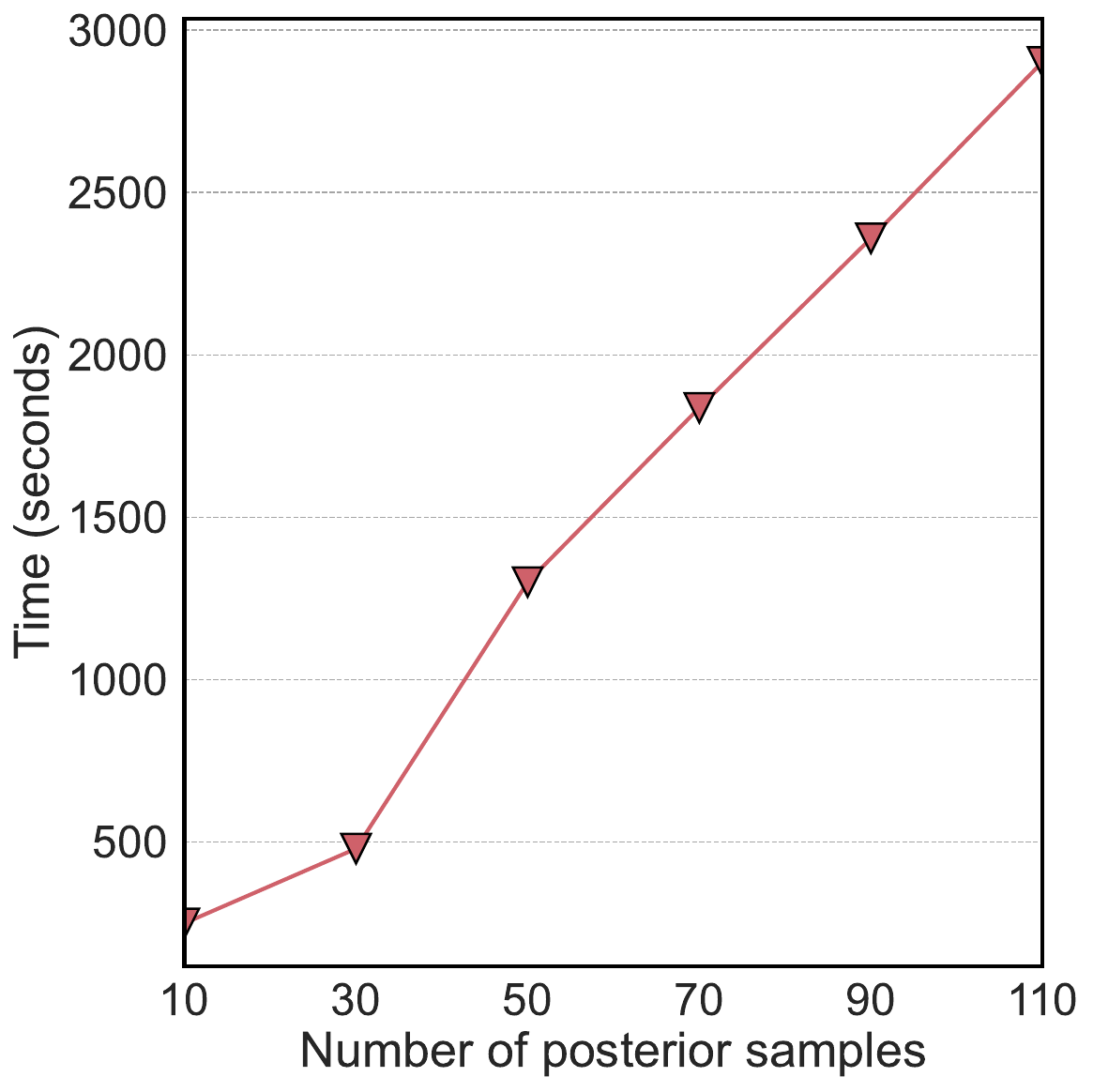}
  \captionsetup{type=figure}
  \vspace{-4mm}
  \caption{Runtime vs.\ \nsamples\ at 30\% \mcar\ on \wine\ for \tabcsdis.}
  \label{fig:sampling-time-S-variants-tab}  
\end{minipage}
\hfill
\begin{minipage}[t]{0.48\columnwidth}
  \centering
\includegraphics[width=\linewidth]{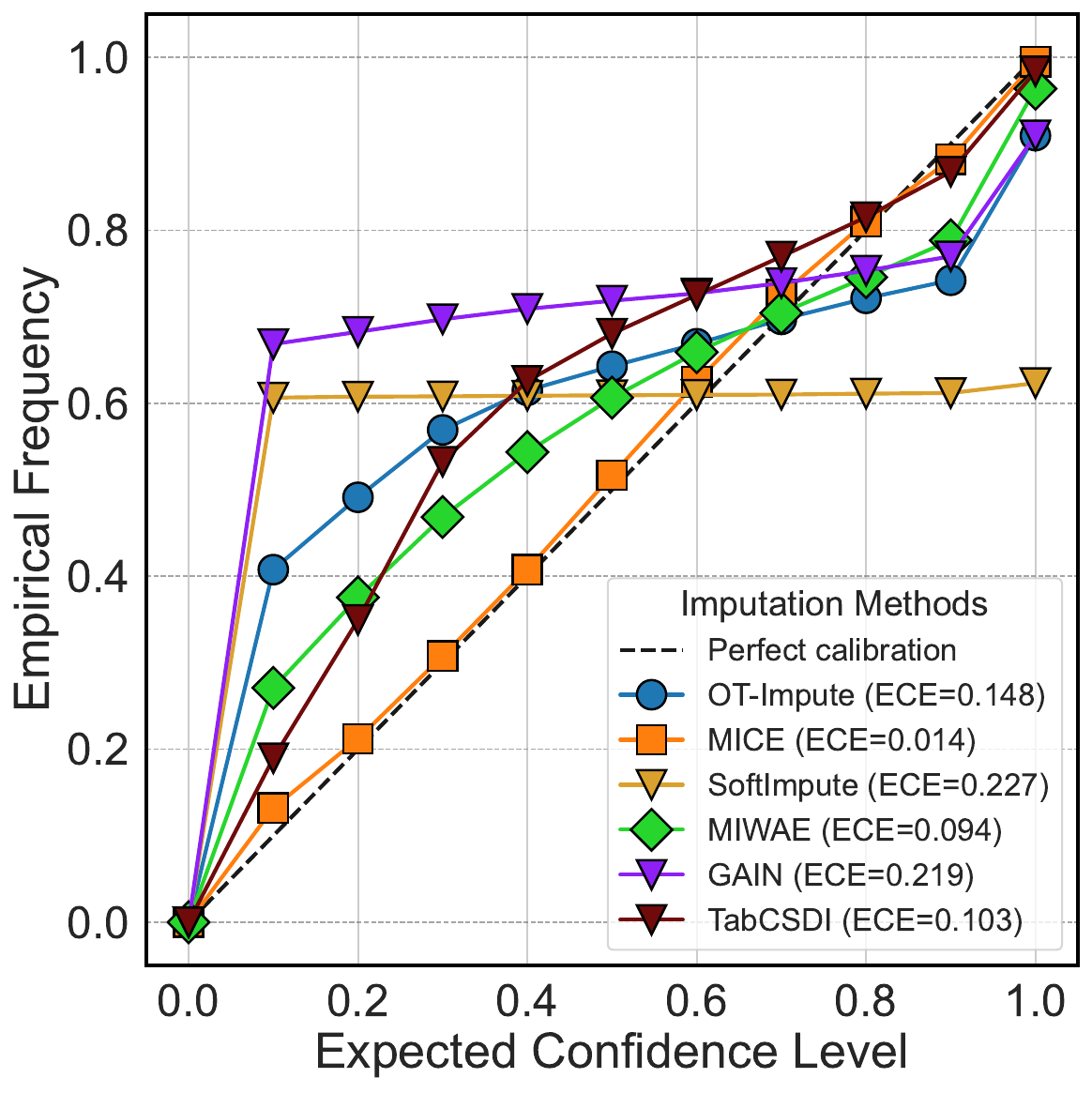}
  \captionsetup{type=figure}
  \vspace{-4mm}
  \caption{Calibration curves for 30\% \mcar\ on \qsar.}
  \label{fig:cal-all-added}
\end{minipage}

\end{center}

\subsection{Runtime per Method}\label{sec:runtime}
Classical imputers such as \mice, \ot, and \softimpute operate directly on masked data without training, so their total runtime corresponds to iterative convergence. Learning-based models (\gain, \miwae, \tabcsdi) include both training and inference time.

\mice\ is generally faster than other methods in Figure~\ref{fig:timebar-wine} because it iteratively fits simple per-feature regressions on the currently observed data rather than training complex models or performing extensive sampling. However, in \housing, the large number of attributes increases the per-iteration workload (more predictors per regression). \ot involves repeated Sinkhorn computations over many pairs and iterations, which can be moderately costly, yet it remains faster than deep learning and diffusion models since it avoids lengthy neural-network training and sampling. Its runtime also appears nearly constant across datasets because a fixed batch size and iteration count are used. \softimpute\ is slower than \mice\ and can even be slower than \ot\ and deep learning models on larger datasets due to repeated SVD operations. Among the generative models, \gain\ is generally faster than \miwae\ because it avoids multiple decoder evaluations per sample. Diffusion-based \tabcsdi\ is the most computationally expensive method, as both training and inference require numerous sequential denoising steps.

For uncertainty estimation, running five independent imputations increases the total runtime roughly fivefold. The sampling-based variants \miwaes and \gains are faster overall because they train once and generate multiple imputations from the same model; however, sampling increases inference time which makes \texttt{-S} slower than \texttt{-U}. Diffusion-based \tabcsdis remains the most time-consuming due to its repeated reverse chains during sampling. The plots in Figure~\ref{fig:sampling-time-S-variants} and Figure~\ref{fig:sampling-time-S-variants-tab} show runtime versus the number of posterior samples for the \texttt{S}-variant methods. As expected, taking more posterior samples increases runtime roughly linearly, since inference repeats the sampling loop more times.

\subsection{Imputation Accuracy}
\label{sec:acc}

\begin{table*}[h]
\centering
\setlength{\tabcolsep}{4pt}
\renewcommand{\arraystretch}{1.1}
\begin{tabular}{llcccccc}
\toprule
\multicolumn{2}{c}{\textbf{Dataset / Mechanism}} & \mice & \ot & \softimpute & \miwae & \gain & \tabcsdi \\
\midrule
\multirow{3}{*}\wine & \mcar & 0.784 $\pm$ 0.071 & \bestcell{$0.570 \pm 0.004$} & \secondcell{$0.603 \pm 0.003$} & 0.567 $\pm$ 0.016 & 0.788 $\pm$ 0.010 & 1.046 $\pm$ 0.033 \\
 & \mar & 0.801 $\pm$ 0.073 & \bestcell{$0.628 \pm 0.004$} & \secondcell{$0.639 \pm 0.016$} & 0.582 $\pm$ 0.015 & 0.800 $\pm$ 0.010 & 1.103 $\pm$ 0.042 \\
 & \mnar & 0.818 $\pm$ 0.084 & \bestcell{$0.655 \pm 0.003$} & \secondcell{$0.587 \pm 0.001$} & 0.621 $\pm$ 0.015 & 0.807 $\pm$ 0.014 & 1.076 $\pm$ 0.021 \\
\midrule

\multirow{3}{*}\energy & \mcar & 0.541 $\pm$ 0.044 & 0.553 $\pm$ 0.005 & \secondcell{$0.482 \pm 0.009$} & \bestcell{$0.431 \pm 0.008$} & 0.803 $\pm$ 0.011 & 0.958 $\pm$ 0.037 \\
 & \mar & 0.579 $\pm$ 0.046 & 0.619 $\pm$ 0.006 & \secondcell{$0.472 \pm 0.003$} & \bestcell{$0.425 \pm 0.010$} & 0.789 $\pm$ 0.009 & 0.977 $\pm$ 0.032 \\
 & \mnar & 0.657 $\pm$ 0.039 & 0.796 $\pm$ 0.006 & \secondcell{$0.543 \pm 0.000$} & \bestcell{$0.578 \pm 0.011$} & 0.868 $\pm$ 0.016 & 1.151 $\pm$ 0.047 \\
\midrule
\multirow{3}{*} \housing & \mcar & 0.779 $\pm$ 0.003 & 0.568 $\pm$ 0.001 & \secondcell{$0.472 \pm$ 0.003} & \bestcell{$0.406 \pm 0.005$} & 0.654 $\pm$ 0.011 & -- \\
 & \mar & 0.763 $\pm$ 0.004 & 0.499 $\pm$ 0.001 & \secondcell{$0.403 \pm 0.001$} & \bestcell{$0.306 \pm 0.003$} & 0.546 $\pm$ 0.011 & -- \\
 & \mnar & 0.757 $\pm$ 0.001 & 0.553 & \secondcell{$0.482$} & \bestcell{$0.391 \pm 0.004$} & 0.624 $\pm$ 0.027 & -- \\
\midrule
\multirow{3}{*}\qsar & \mcar & 0.3307 $\pm$ 0.0002 & \bestcell{$0.3611 \pm 0.0020$} & \secondcell{$0.4249 \pm 0.0174$} & 0.5145 $\pm$ 0.0661 & 0.5439 $\pm$ 0.0044 & 0.9137 $\pm$ 0.1421 \\
 & \mar & 0.3406 $\pm$ 0.0023 & \bestcell{$0.3798 \pm 0.0021$} & \secondcell{$0.4533 \pm 0.0151$} & 0.5286 $\pm$ 0.0625 & 0.5393 $\pm$ 0.0047 &  - \\
 & \mnar & 0.5243 $\pm$ 0.0090 & \bestcell{$0.6780 \pm 0.0015$} & \secondcell{$0.7377 \pm 0.0264$} & 0.5959 $\pm$ 0.0464 & 0.6606 $\pm$ 0.0118 &  - \\
\midrule
\multirow{3}{*} \cancer & \mcar & 0.911 $\pm$ 0.002 & 0.691 & \secondcell{$0.607$} & \bestcell{$0.497 \pm 0.010$} & 0.735 $\pm$ 0.016 & -- \\
 & \mar & 0.966 $\pm$ 0.002 & 0.705 & \secondcell{$0.601$} & \bestcell{$0.506 \pm 0.009$} & 0.662 $\pm$ 0.011 & -- \\
 & \mnar & 0.964 $\pm$ 0.001 & 0.824 & \secondcell{$0.632$} & \bestcell{$0.528 \pm 0.014$} & 0.817 $\pm$ 0.036 & -- \\
\bottomrule
\end{tabular}%
\caption{\mae\ at 30\% missingness (mean $\pm$ std).}
\label{tab:mae30_core}
\end{table*}

\begin{table}[t]
\centering
\setlength{\tabcolsep}{2.5pt}      
\renewcommand{\arraystretch}{1.05} 
\resizebox{\columnwidth}{!}{       
\begin{tabular}{llccccc}
\toprule
\multicolumn{2}{c}{\textbf{Dataset/Mech.}} 
& \miwaes & \miwaeu & \gains & \gainu & \tabcsdis \\ 
\midrule
\multirow{3}{*}{\wine} 
 & \mcar & \bestcell{0.568} & \secondcell{0.587} & 0.773 & 0.797 & 0.805 \\
 & \mar  & \bestcell{0.603} & \secondcell{0.601} & 0.781 & 0.791 & 0.868 \\
 & \mnar & \bestcell{0.620} & \secondcell{0.643} & 0.790 & 0.826 & 0.890 \\
\midrule
\multirow{3}{*}{\energy} 
 & \mcar & \bestcell{0.434} & \secondcell{0.462} & 0.770 & 0.769 & 0.883 \\
 & \mar  & \bestcell{0.434} & \secondcell{0.443} & 0.771 & 0.742 & 0.919 \\
 & \mnar & \bestcell{0.585} & \secondcell{0.608} & 0.867 & 0.797 & 1.107 \\
\midrule
\multirow{3}{*}{\housing} 
 & \mcar & 0.412 & \bestcell{0.404} & 0.660 & \secondcell{0.603} & -- \\
 & \mar  & 0.311 & \bestcell{0.308} & 0.529 & \secondcell{0.522} & -- \\
 & \mnar & 0.397 & \bestcell{0.385} & 0.605 & \secondcell{0.597} & -- \\
\midrule
\multirow{3}{*}{\qsar} 
 & \mcar & 0.4235 & \bestcell{0.4135} & \secondcell{0.5445} & 0.5136 & 0.6835 \\
 & \mar  & \secondcell{0.4515} & \bestcell{0.4319} & 0.5362 & 0.4983 & -- \\
 & \mnar & \bestcell{0.7360} & 0.7421 & 0.6550 & \secondcell{0.6092} & -- \\
\midrule
\multirow{3}{*}{\cancer} 
 & \mcar & \secondcell{0.523} & \bestcell{0.515} & 0.705 & 0.693 & -- \\
 & \mar  & \secondcell{0.537} & \bestcell{0.501} & 0.649 & 0.688 & -- \\
 & \mnar & \bestcell{0.515} & \secondcell{0.541} & 0.845 & 0.730 & -- \\
\bottomrule
\end{tabular}
}
\caption{\mae\ at 30\% missingness for \texttt{-S}/\texttt{-U} variants.}
\label{tab:mae30_variants}
\end{table}

\begin{figure}[h]
  \centering
  \begin{subfigure}[t]{0.24\textwidth}
    \includegraphics[width=\linewidth]{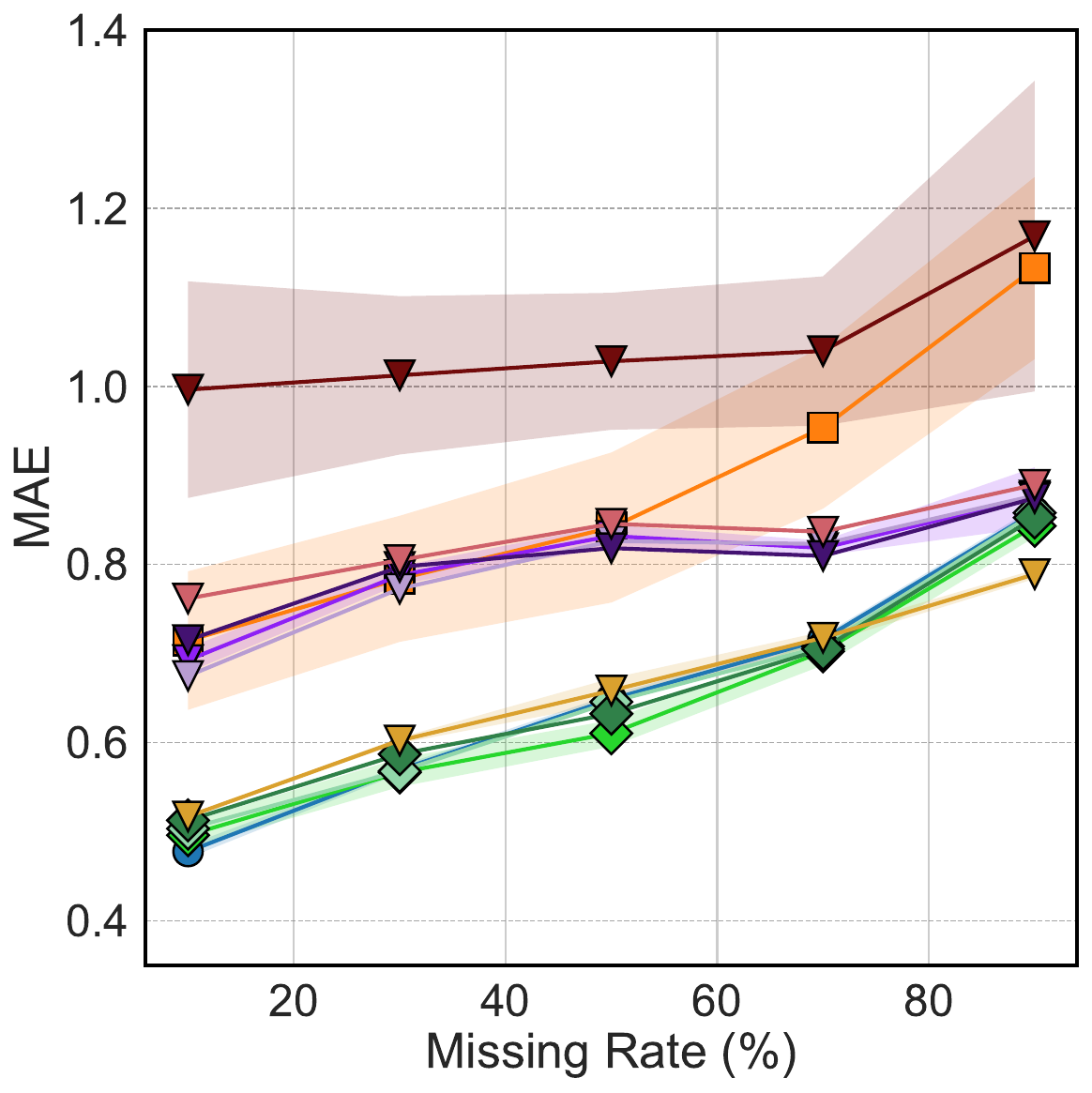}
    \caption{\wine}
    \label{fig:wine-mae-mcar}
  \end{subfigure}
  \begin{subfigure}[t]{0.24\textwidth}
    \includegraphics[width=\linewidth]{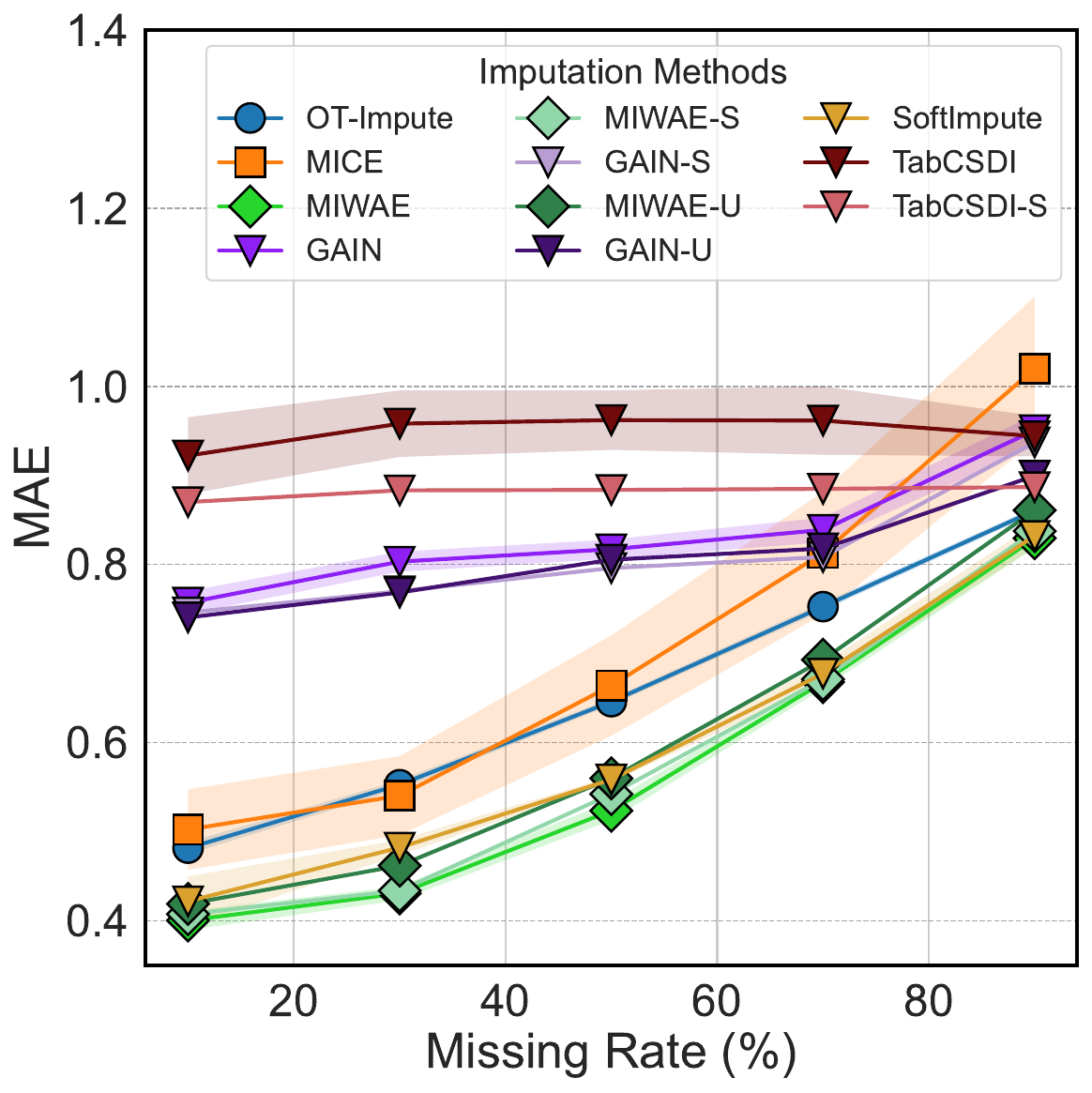}
    \caption{\energy}
    \label{fig:energy-mae-mcar}
  \end{subfigure}\hfill
  \caption{\mae\ vs.\ missing rate for \mcar.}
  \label{fig:mae-vs-missingrate}
\end{figure}
Figure~\ref{fig:mae-vs-missingrate} shows the \mae\ across varying missingness rates for the \wine\ and \energy\ datasets. As expected, higher missingness consistently leads to higher \mae\ across all methods, because with fewer observed entries, models have less information to infer dependencies between attributes, reducing the accuracy of the reconstructed values.

At 30\% missingness (Tables~\ref{tab:mae30_core}–\ref{tab:mae30_variants}), \mae\ generally increases from \mcar\ to \mnar. Exceptions appear in \housing, where strong feature collinearity allows \mar/\mnar\ imputations to outperform \mcar. \miwae achieves the best accuracy because its multiple-sampling and importance-weighting approach allows it to better approximate the true data distribution and produce more reliable imputations. Another key observation is that for datasets like \wine and \qsar the {\ot} performs best due to its effective distribution-matching.
\softimpute often ranks second best overall, where its nuclear-norm regularization effectively captures latent structure.

\subsection{Calibration Curves and ECE} \label{sec:uncert-results}

\begin{figure}
\centering
\graphicspath{{figs/}} 
\begin{subfigure}{0.24\textwidth}
\includegraphics[width=\linewidth]{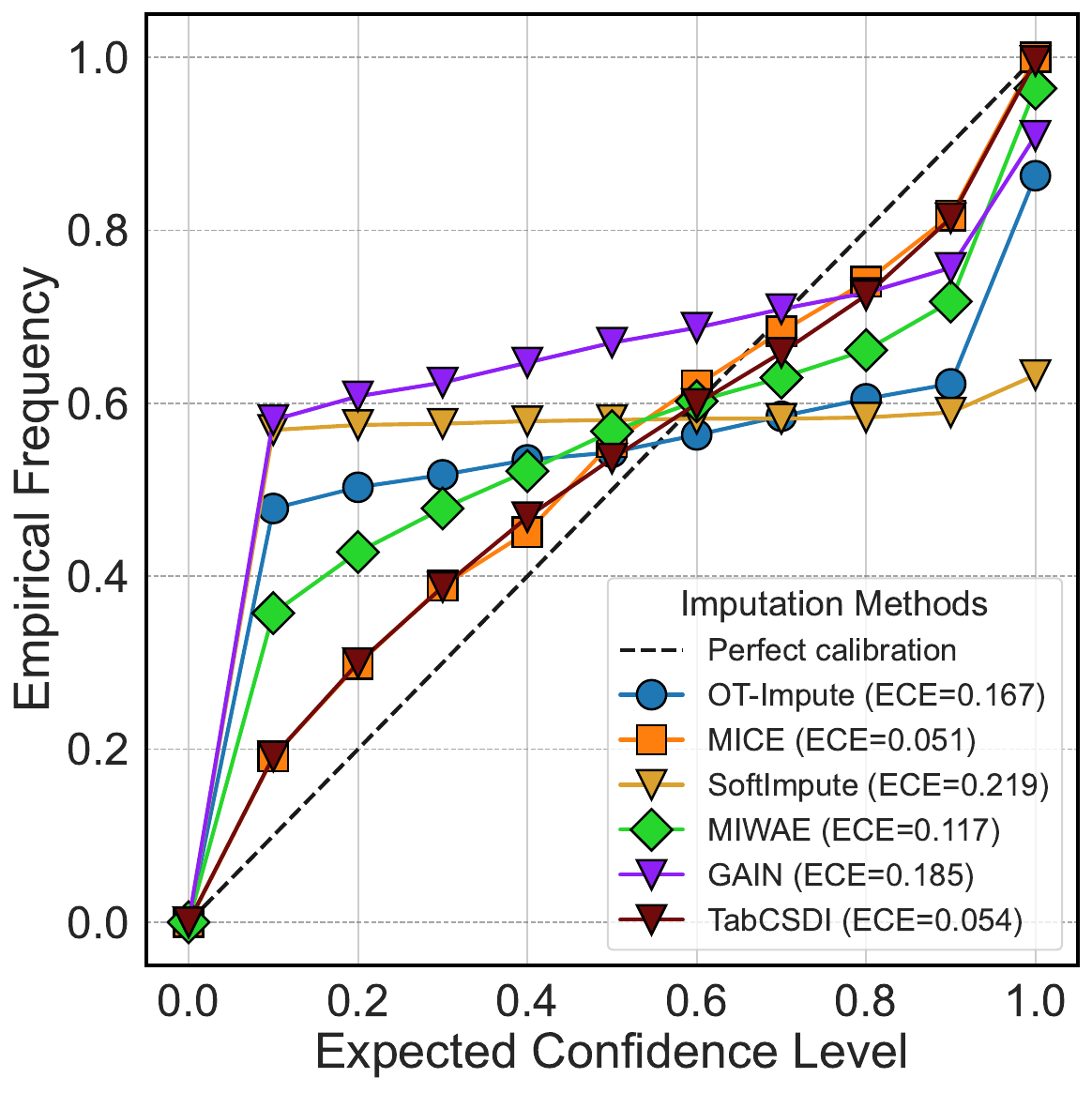}
    \caption{\wine}
    \label{fig:cal-all-wine}
\end{subfigure}
\hfill
\begin{subfigure}{0.24\textwidth}    \includegraphics[width=\linewidth]{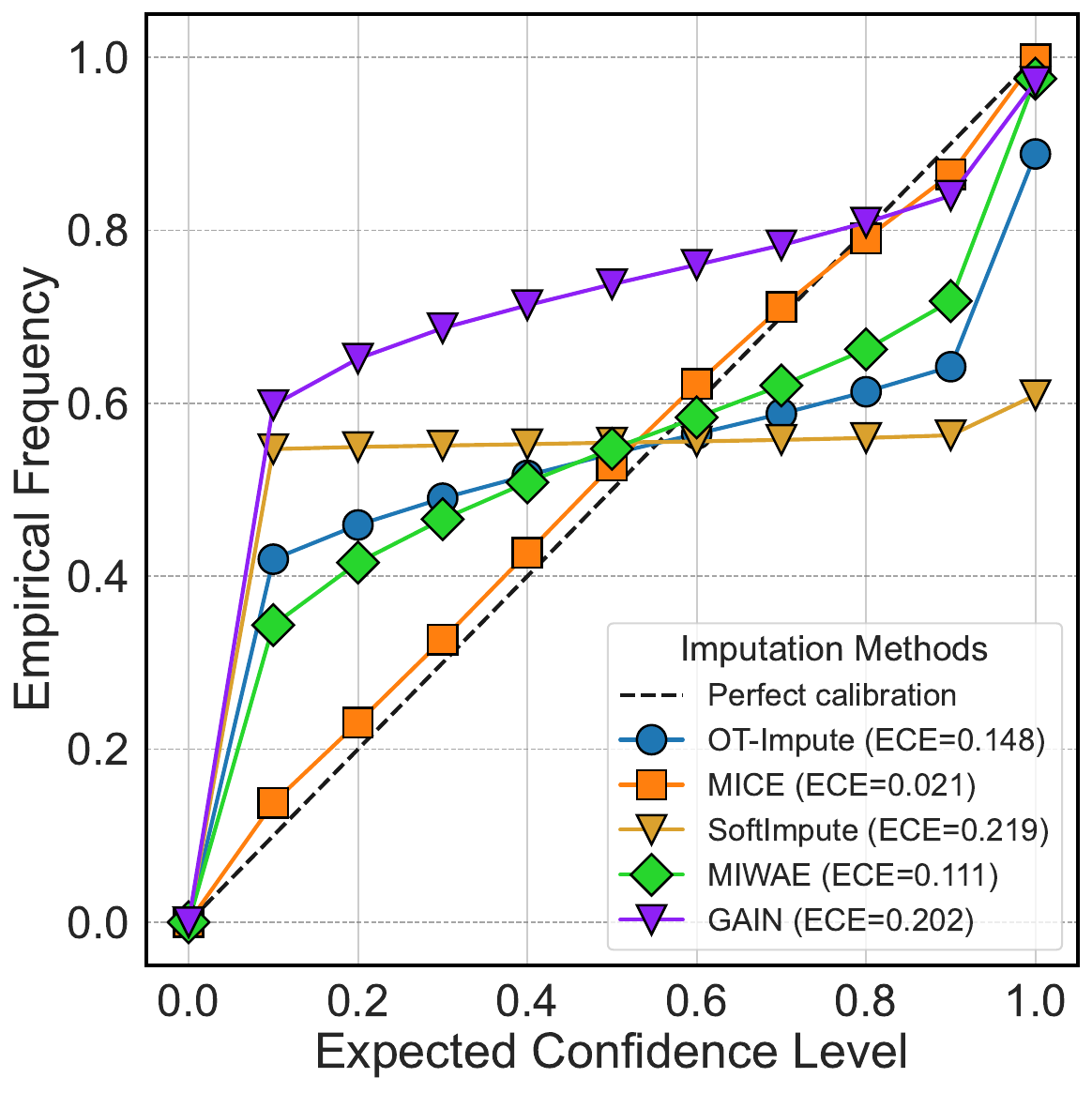}
    \caption{\housing}
    \label{fig:cal-all-housing}
\end{subfigure}
\caption{Calibration curves  for 30\% \mcar.}
\label{fig:cal-all}
\end{figure}


Figures~\ref{fig:cal-all-added} and~\ref{fig:cal-all} present calibration curves across imputation methods and datasets at a 30\% \mcar. 
Overall, \mice\ achieves the most reliable calibration, producing curves near the ideal diagonal and the lowest \ece\ by incorporating realistic residual noise and averaging variability across multiple runs. 
Across datasets and mechanisms, \softimpute\ is the least calibrated. As a deterministic low-rank method, it provides only point estimates without modeling uncertainty. Post-hoc proxies yield overly narrow, flat calibration curves, reflecting constant and unreliable uncertainty across confidence levels.
Unlike \mice\ or generative models, \ot\ lacks per-cell predictive distributions and instead aligns global feature distributions. This yields accurate point imputations but poorly sized uncertainty ranges, leading to higher \ece\ and mixed calibration—over-confident when transport mass is concentrated and under-confident when dispersed. Still, it calibrates slightly better than \softimpute\  and \gain due to modest stochasticity from mini-batching.

\begin{figure}
\centering
\graphicspath{{figs/}} 
\begin{subfigure}{0.24\textwidth}
    \includegraphics[width=\linewidth]{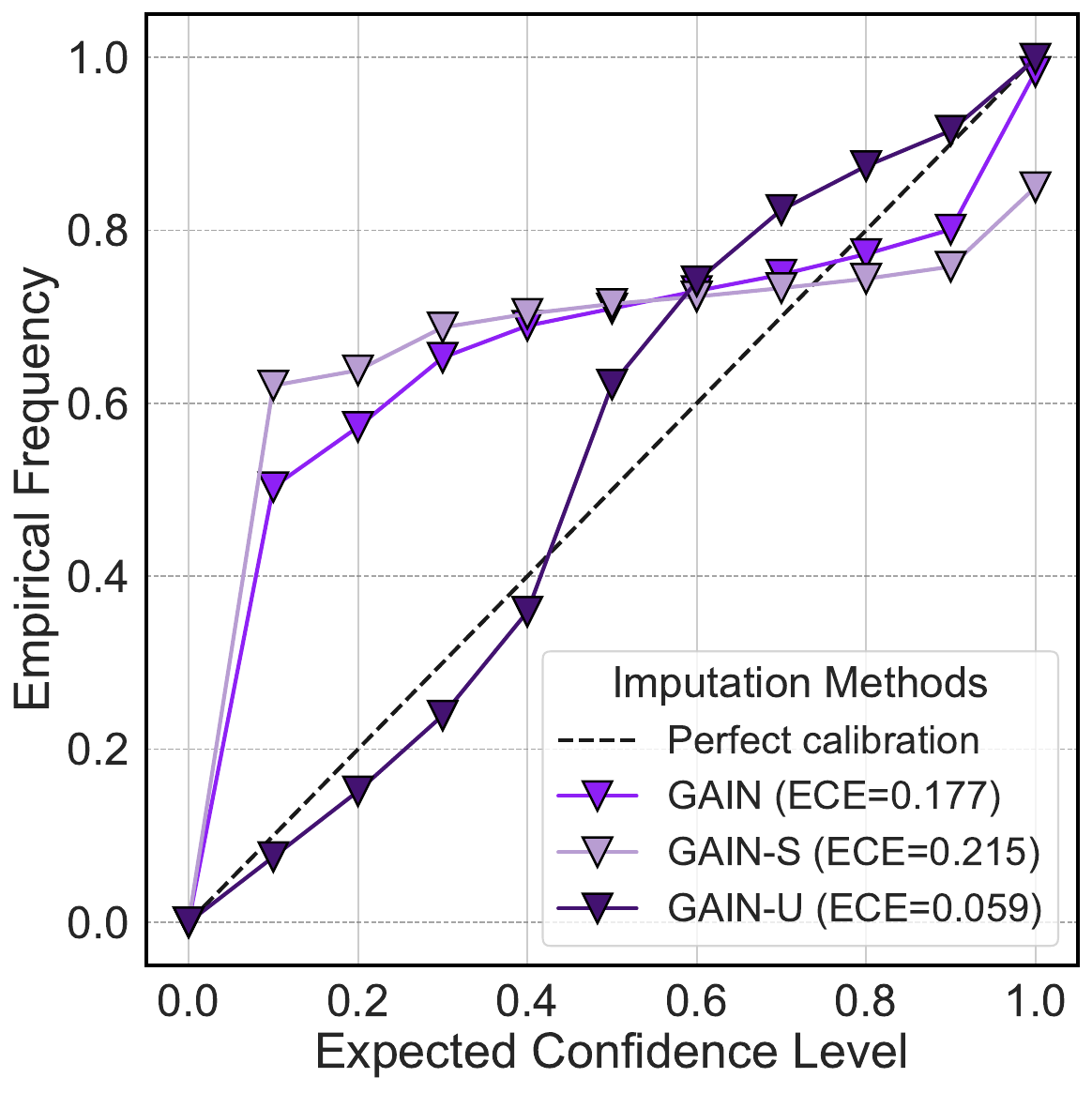}
    \caption{\cancer}
    \label{fig:cal-gain-cancer}
\end{subfigure}
\begin{subfigure}{0.24\textwidth}
    \includegraphics[width=\linewidth]{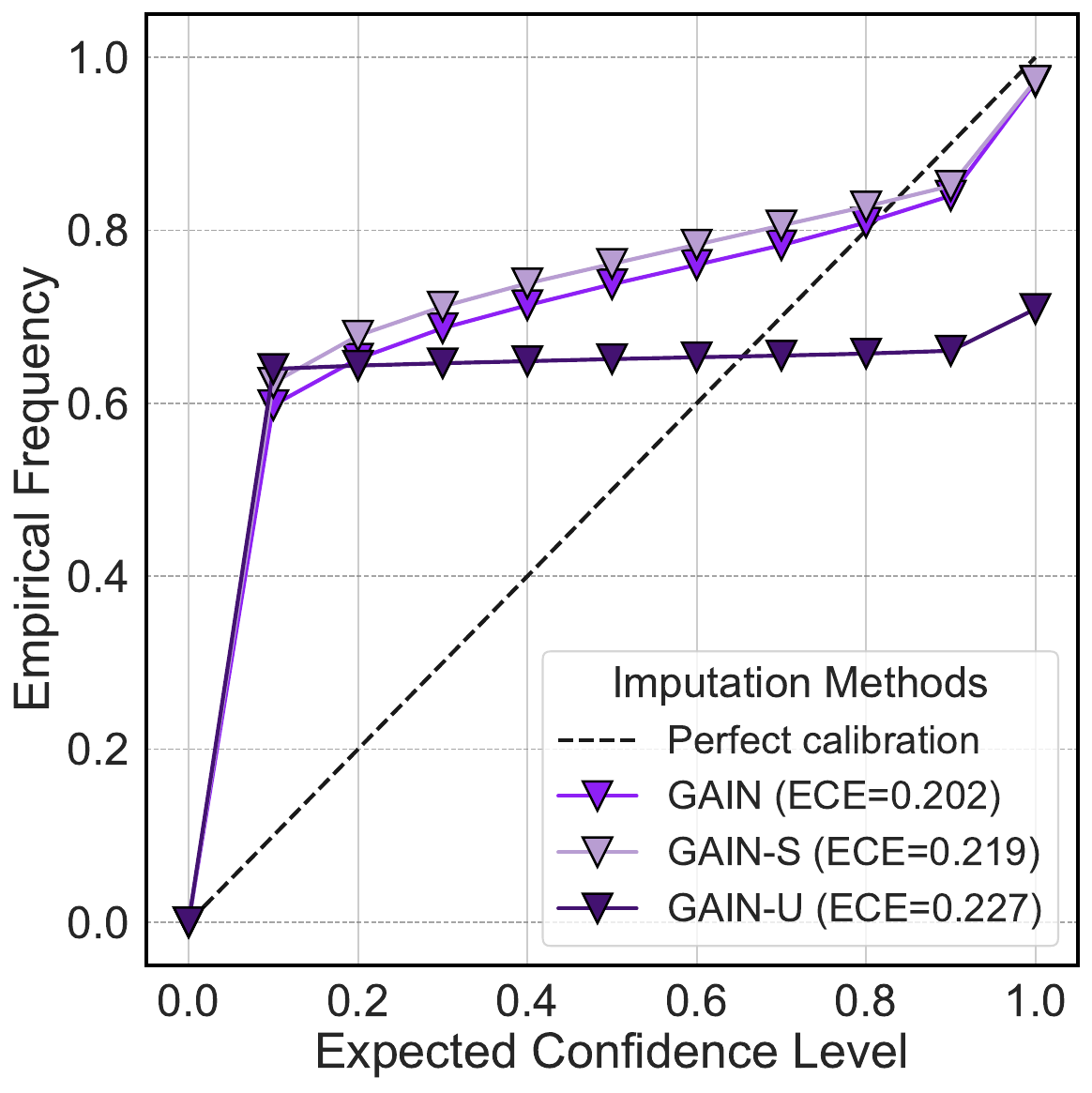}
    \caption{\housing}
    \label{fig:cal-gain-housing}
\end{subfigure}
\caption{Calibration curves for \gain family in 30\% \mcar.}
\label{fig:cal-all-gain}
\end{figure}
\gain\ focuses on producing realistic imputations for the discriminator rather than calibrated uncertainty, leading to over-dispersed (under-confident) predictions and higher \ece\ (Figure~\ref{fig:cal-all-gain}). The \gainu\ variant introduces a variance head to estimate per-cell mean and variance, enabling multiple imputations and improving calibration in some datasets like \cancer\ (Figure~\ref{fig:cal-gain-cancer}), though the added complexity can cause unstable training and noisy variance estimates, as seen in \housing\ (Figure~\ref{fig:cal-gain-housing}). The sampling-only \gains\ variant, which generates multiple noisy samples without learning variance, performs slightly worse than \gain\ because averaging samples tends to smooth out meaningful variability.

\begin{figure}
\centering
\graphicspath{{figs/}} 
\begin{subfigure}{0.24\textwidth}
    \includegraphics[width=\linewidth]{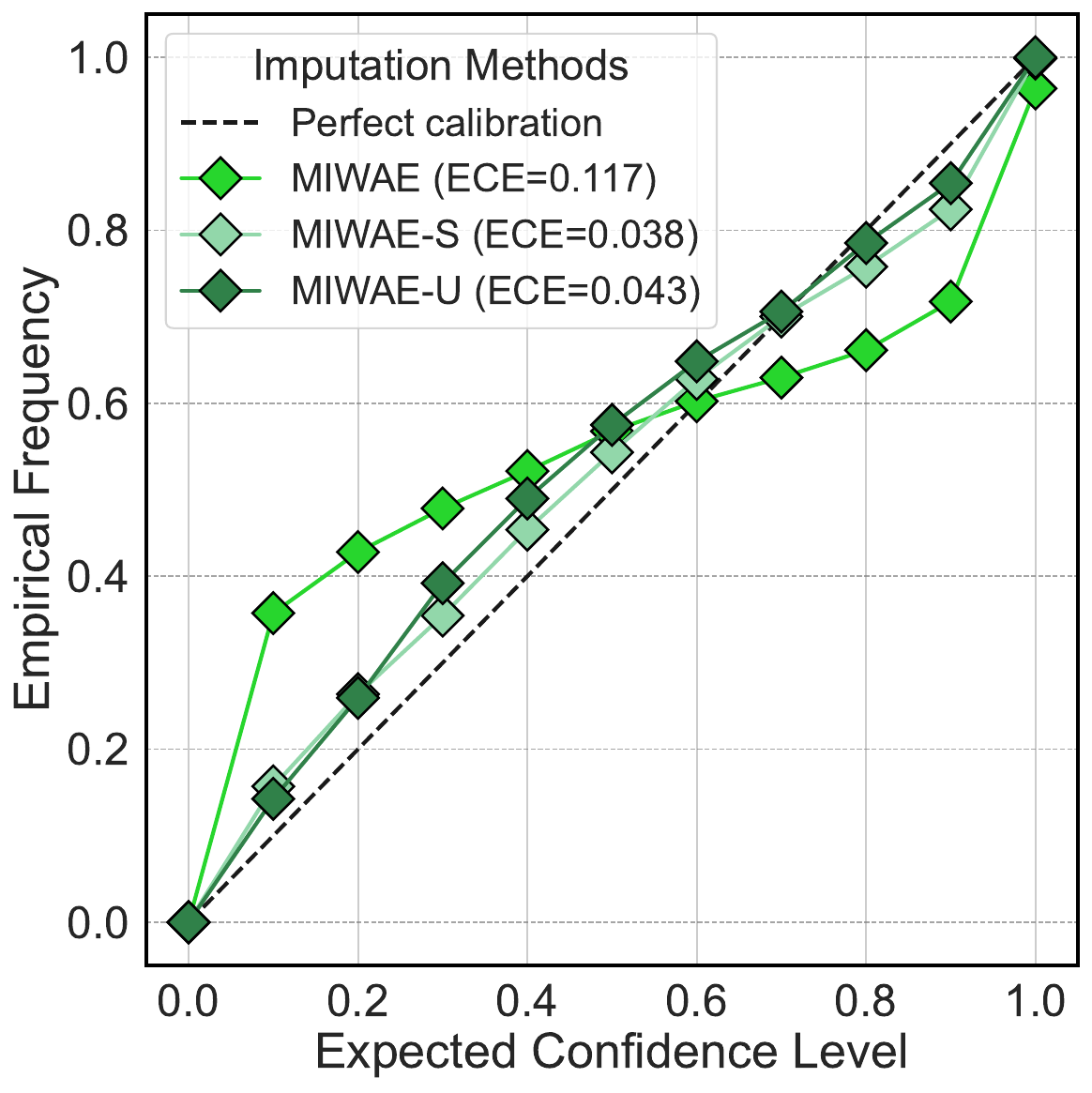}
    \caption{\wine}
    \label{fig:cal-miwae-wine}
\end{subfigure}
\begin{subfigure}{0.24\textwidth}
    \includegraphics[width=\linewidth]{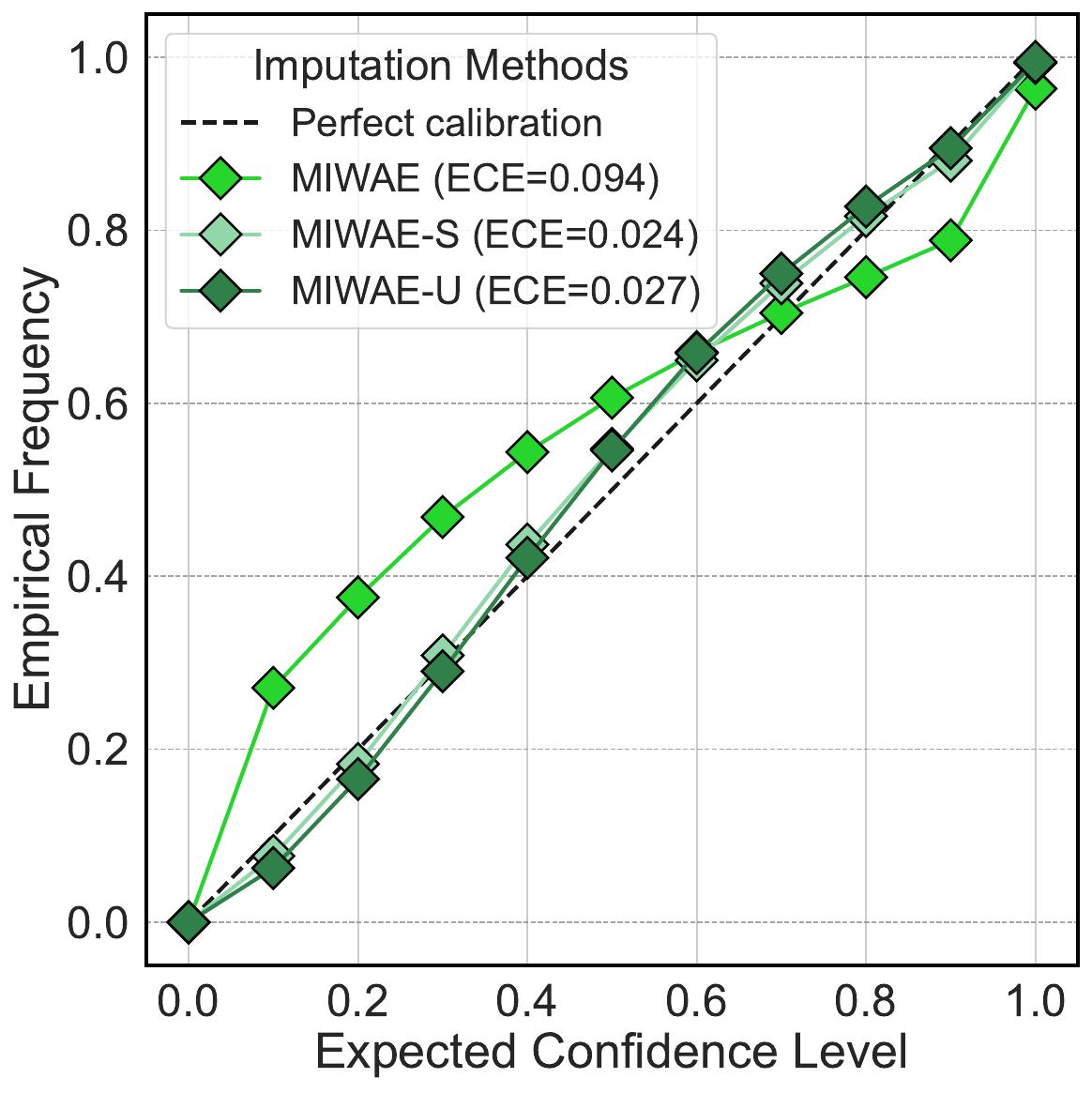}
    \caption{\qsar}
    \label{fig:cal-miwae-qsar}
\end{subfigure}
\caption{Calibration curves for \miwae family in 30\% \mcar.}
\label{fig:cal-all-miwae}
\end{figure}

\miwae\ is generally well-calibrated, it models each missing cell with a full predictive distribution (mean and variance) and averages multiple imputations at inference. This stochastic approach captures genuine variability in the data, preventing over-tight or overly diffuse predictions and thus lowering \ece. Its variants, \miwaes\ and \miwaeu in Figure~\ref{fig:cal-all-miwae}, further refine the predictive spread, \miwaes\ by adjusting variance to correct over/under-confidence, and \miwaeu\ by directly using model-predicted uncertainty, resulting in confidence intervals that better match empirical coverage and occasionally outperform \mice (comparison between Figures~\ref{fig:cal-all-wine} and ~\ref{fig:cal-miwae-wine})

\begin{figure}
\centering
\graphicspath{{figs/}} 
\begin{subfigure}{0.24\textwidth}
    \includegraphics[width=\linewidth]{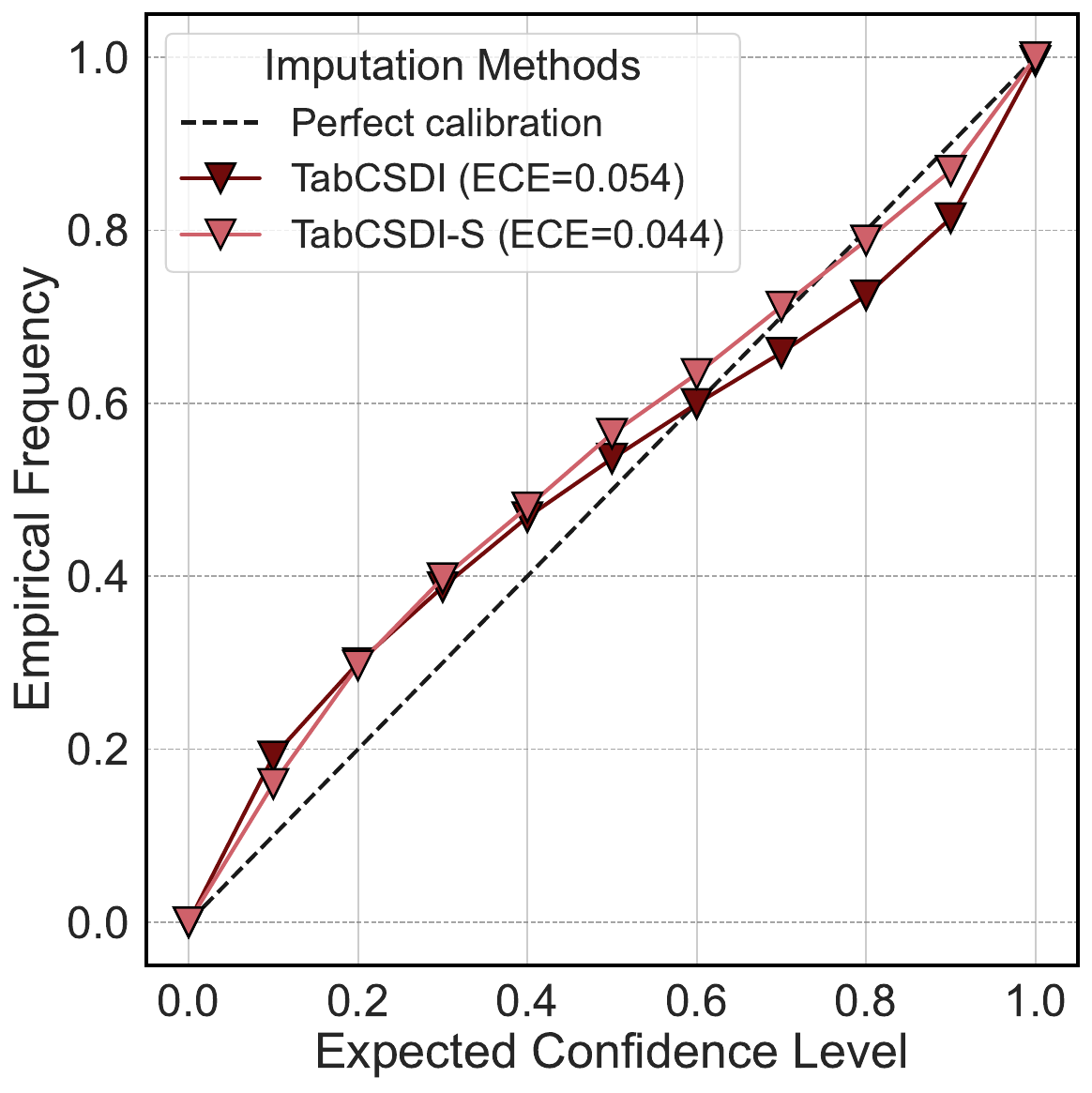}
    \caption{\wine}
    \label{fig:cal-tabcsdi-wine}
\end{subfigure}
\begin{subfigure}{0.24\textwidth}
    \includegraphics[width=\linewidth]{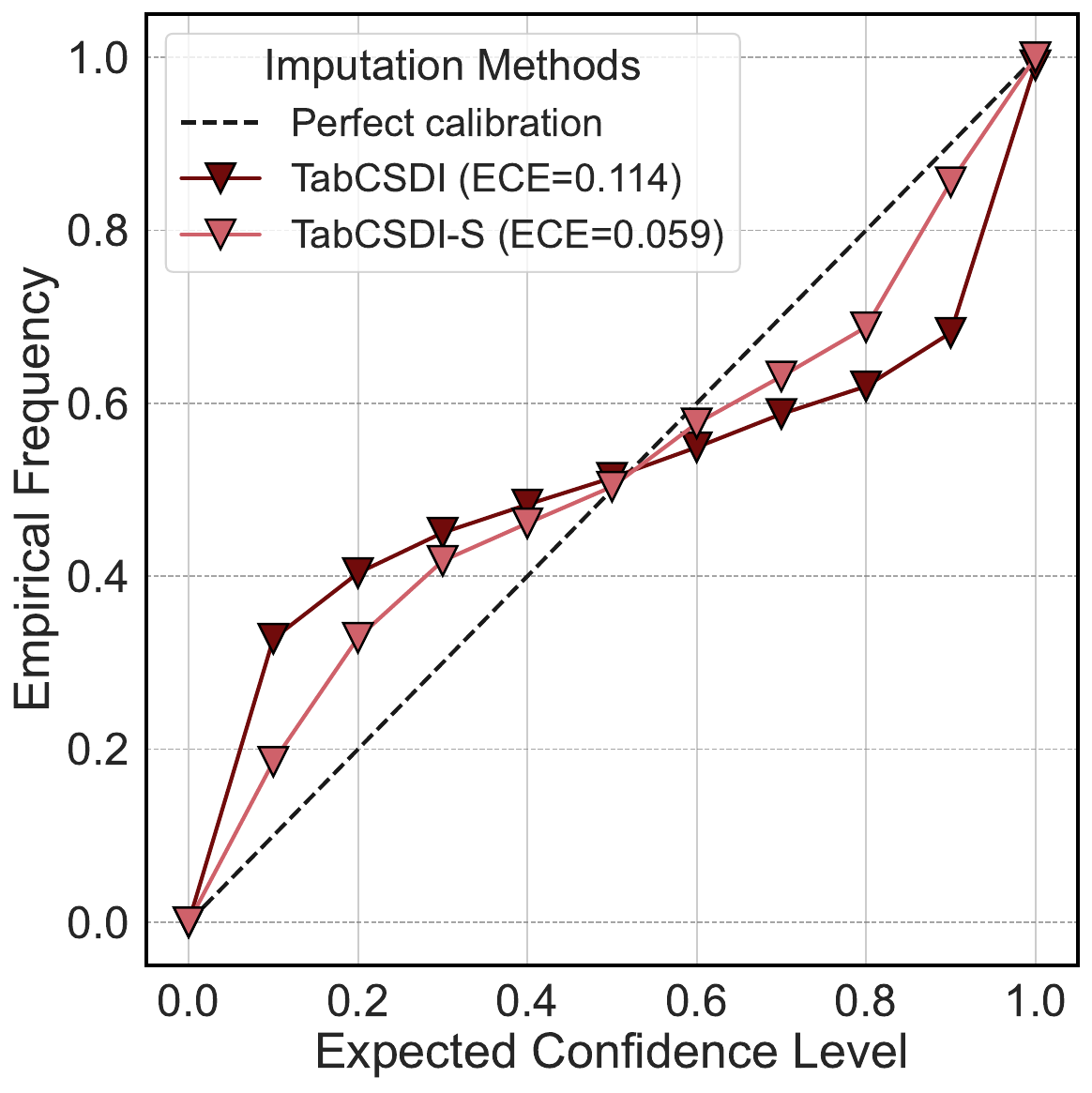}
    \caption{\energy}
    \label{fig:cal-tabcsdi-energy}
\end{subfigure}
\caption{Calibration curves for \tabcsdi in 30\% \mcar.}
\label{fig:cal-all-tabcsdi}
\end{figure}

{\tabcsdi} shows low \ece across all mechanisms in {\wine} (Figure~\ref{fig:cal-tabcsdi-wine}) but shows mixed confidence in {\energy} (Figure~\ref{fig:cal-tabcsdi-energy}). In {\wine}, strong feature correlations and limited samples cause wide intervals, while in {\energy}, varying feature difficulty leads to over-confidence on easy attributes and under-confidence on hard ones. The sampling-based {\tabcsdis} further improves calibration by averaging multiple diffusion trajectories, producing more accurate uncertainty estimates and reducing both over- and under-confidence.

\section{Analysis and Takeaways}\label{sec:analysis}

No single method dominates across datasets in accuracy, calibrated uncertainty, or runtime; the choice should weigh these trade-offs. Across mechanisms, we observe a consistent degradation from \mcar\ to \mar\ to \mnar\ in both accuracy and calibration, with a few dataset-specific exceptions. Methods also differ in run-to-run robustness: variance across seeds is generally modest for \mice/\softimpute/\miwae and higher for diffusion-based models (\tabcsdi/\tabcsdis), motivating the reporting of mean~$\pm$~std over multiple runs. In our experiments, \softimpute\ delivers strong accuracy across most datasets but remains consistently poorly calibrated, making it unsuitable when trustworthy uncertainty estimates are needed. \mice\ provides the most dependable calibration overall, though its point accuracy is generally lower than leading alternatives. \miwae\ strikes a favorable middle ground, achieving both good accuracy and reasonably strong calibration, albeit with the highest computational cost. Therefore, the choice of method should depend on the main priority, whether accuracy, calibrated uncertainty, or runtime efficiency, and the characteristics of the dataset.

\section{Conclusion and Future Work}\label{sec:conclusion}

This work evaluated uncertainty \emph{calibration} in imputation rather than accuracy alone. Across six representative methods and multiple datasets, rates, and mechanisms, we quantified uncertainty using three complementary strategies: repeated runs, sampling from a trained model, and direct predictive distributions. Our results show that accuracy and calibration are distinct: methods with strong point error  may be poorly calibrated, while methods that explicitly model uncertainty tend to yield more reliable coverage. In short, imputers should be judged not only by how close they get on average, but by whether their stated confidence matches observed frequencies.

There remains substantial room for future work. Our experiments focused on numeric tabular datasets, and an important next step is to extend the framework to categorical and mixed-type data. Handling heterogeneous attributes requires different forms of uncertainty representation and calibration, and developing a unified approach for such data remains challenging.
Because \tabcsdi is computationally intensive, we were unable to evaluate it on all datasets; completing this analysis is part of future work.
Another direction is to separate aleatoric and epistemic uncertainty and study how each behaves. Simulation can help distinguish them: epistemic uncertainty should decrease with more data or stronger models, while aleatoric uncertainty reflects inherent noise that persists.
A final avenue is to examine how calibrated uncertainty interacts with downstream tasks such as fairness-sensitive prediction, risk assessment, and human-in-the-loop decision making. Integrating calibrated uncertainty into interactive workflows may improve reliability, interpretability, and decision support.

\bibliographystyle{ieeetr}
\bibliography{ref-short}
\end{document}